\RequirePackage{fix-cm}
\documentclass[epjc3]{svjour3}  

\RequirePackage{graphicx,amssymb,amsfonts,amsmath,geometry,hyperref,color,lipsum}
\usepackage[english]{babel} 
\usepackage[T1]{fontenc}
\usepackage{float}
\usepackage{pdflscape}
\usepackage{multirow}
\begin{document}
\title{Dynamics and statefinder analysis of a class of sign-changeable interacting dark energy scenarios}
\author{Fabiola Arévalo\thanksref{e1,addr1}
        \and
       Antonella Cid\thanksref{e2,addr2,addr2b}
}
\thankstext{e1}{e-mail: fabiola.arevalo@umayor.cl}
\thankstext{e2}{e-mail: acidm@ubiobio.cl}

\institute{Núcleo de Matemática, Física y Estadística, Universidad Mayor, Temuco, Chile.\label{addr1} \and Departamento de Física,  Universidad del Bío-Bío, Casilla 5-C, Concepción, Chile. \label{addr2} \and
Centro de Ciencias Exactas, Universidad del B\'io-B\'io, Casilla 447, Chill\'an, Chile.
\label{addr2b}}

\date{Received: date / Accepted: date} 
\maketitle

\begin{abstract}
We revise the dynamical properties of a class of cosmological models where the dark sector interacts through an interacting term that changes sign during evolution. In particular, we obtain the critical points and we investigate the existence and stability conditions for cosmological solutions, describing radiation, matter and dark energy dominated eras. We find that all the studied models admit a stable critical point corresponding to an accelerated phase. We use background data to find the best fit parameters for one of the studied models, resulting an interacting parameter with a definite sign within $1\sigma$ confidence level, consistent with the results of the dynamical system analysis. We also compute the statefinder parameters and plot the $r-q$ and $r-s$ planes, where we observe different trajectories when we vary the interaction parameter for a specific model and when we vary the interacting scenario. We can in this sense distinguish among models, including $\Lambda$CDM.
\end{abstract}

\section{Introduction}
All the available data in cosmology seems to indicate that our current universe consist of $69\%$ of dark energy and $26\%$ of dark matter \cite{Planck:2018vyg}, but to the present understanding of physics, both components have unknown nature \cite{Weinberg:1988cp,Lin:2019uvt}. Usually we consider the standard cosmological scenario $\Lambda$CDM, where the dark components evolve independently. A different approach is to consider that both components interact \cite{coincidence,interaction}. Interacting scenarios allow to alleviate the coincidence problem \cite{coincidence} and also are useful in addressing the tension in the $H_0$ parameter \cite{H0}.
Among interacting scenarios it is common to consider those models in which the interaction sign remains constant during evolution but there are some less explored scenarios where the interaction sign can change \cite{Wei:2010cs,Arevalo:2019axj,Pan:2019jqh}.

In cosmology, dynamical system methods are a very useful technique to investigate the asymptotic behavior of a given cosmological model, specially when the analytical solutions are unknown. In Ref. \cite{Copeland:2006wr} the authors review the dynamics of a plethora of dark energy models applying dynamical system analysis, an updated version of this work is presented in Ref. \cite{Bahamonde:2017ize}. Both works include interacting scenarios.

The authors of Ref. \cite{Sahni:2002fz} introduce the cosmological pair $r$ and $s$, dubbed statefinder parameters, depending on derivatives of the scale factor up to third order, where the evolution of these parameters allows to differentiate among dark energy models. For instance, in Ref. \cite{Zimdahl:2003wg} the authors use the statefinder parameters to characterize interacting quintessence models of dark energy.

In Refs. \cite{Panotopoulos:2020kpo,Huang:2021zgj} the authors apply dynamical systems methods and perform a statefinder analysis to interacting dark energy models. Even though the dynamical system analysis shows a similar asymptotic behavior for these models, the statefinder diagnosis allows to distinguish among models, when the evolution of the statefinder parameters is studied.

The goal of this work is to revise the dynamics of a class of sign-changeable scenarios which do not have analytical solutions, then it is interesting to characterize the asymptotic behavior of these models through dynamical system analysis and study its performance in fitting the known asymptotic behavior of our universe. The studied models are then compared through the evolution of the statefinder parameters to distinguish its behavior from $\Lambda$CDM.

This work is organized as follows. In section \ref{interaction} we describe interacting scenarios and the class of sign-changeable interactions to study. Section \ref{DS1} is devoted to the dynamical system analysis of each of the models described in section \ref{interaction}. In section \ref{data} we use background data to perform the observational contrast for one of the studied models. In section \ref{SFP} we compute the statefinder parameters and we study the features of the trajectories in the $r-q$ and $r-s$ planes. Finally, in section \ref{FR} we present our final remarks on this work.

\section{Sign-changeable interacting dark energy models}\label{interaction}
Let us consider a flat, homogeneous and isotropic universe in the framework of General Relativity and a cosmological scenario containing radiation ($r$), baryons ($b$), cold dark matter ($c$) and a dark energy ($x$). In the context of an interacting dark energy scenario the conservation equation can be separated into the following equations:
\begin{eqnarray}
\rho'_b+3\rho_b&=&0,\label{4FluidosEC1'}\\
\rho'_r+4\rho_r &=&0,\label{4FluidosEC2'}\\
\rho'_c+3\rho_c &=&-3\Gamma,\label{4FluidosEC3'}\\
\rho'_x&=&3\Gamma,\label{4FluidosEC4'}
\end{eqnarray}
where by convenience we use $\eta=\ln a$ as time variable and we define $(\,)'= d/d\eta$ (where $a$ is the scale factor).
Notice that $\Gamma > 0$ indicates an energy transfer from dark matter to dark energy and $\Gamma < 0$ indicates the opposite.

In this work we study the following sign-changeable linear interactions,
\begin{eqnarray}
\Gamma_{1i}&=&\alpha_{1i}\, q\, \rho_i,\qquad i=T,d,c,x\label{Int_i}\\
\Gamma_{2j}&=&\alpha_{2j}\, q\, \rho'_j,\qquad j=T,d,c\label{Int_j}
\end{eqnarray}
where $\alpha_{1i}$ and $\alpha_{2j}$  are constants, $q$ is the deceleration parameter defined as $q=-1-H'/H$ and $H$ is the Hubble expansion rate. The subscripts $T$ and $d$ denote total energy density and dark sector energy density, ($\rho_c+\rho_x$), respectively.
It was noticed in Ref.\cite{Arevalo:2016epc} that interaction $\Gamma_{1T}$ has analytical solutions for the dark sector energy densities, however the other scenarios do not, to the authors knowledge.  We do not include $\Gamma_{2x}$ in our analysis given that this model leads to a constant deceleration parameter in an interacting scenario. 
Notice that all the studied models reduces to the $\Lambda$CDM scenario when the interacting parameter $\alpha_{1i}$ or $\alpha_{2j}$ is set to zero.

\section{Dynamical System Analysis}\label{DS1} 

In this section we apply dynamical system methods \cite{Copeland:2006wr,Bahamonde:2017ize} in order to identify the relevant cosmological eras in the studied models.
In this sense, we rewrite the system of Eqs. \eqref{4FluidosEC1'}-\eqref{4FluidosEC4'} in terms of the density parameters $\Omega_i=\rho_i/3H^2$ (dimensionless) and we use the Friedmann equation as a constraint among parameters, i.e.,
\begin{equation}
\Omega_r+\Omega_b+\Omega_c+\Omega_x=1,\label{const0}
\end{equation}
to reduce  the system of equations \eqref{4FluidosEC1'}-\eqref{4FluidosEC4'} to:
\begin{eqnarray}
\Omega_r'&=&\Omega_r(-1+\Omega_r-3\Omega_x),\label{eq1}\\
\Omega_c'&=&-\frac{\Gamma}{H^2}+\Omega_c(\Omega_r-3\Omega_x),\label{eq2}\\
\Omega_x'&=&\frac{\Gamma}{H^2}+\Omega_x(3+\Omega_r-3\Omega_x),\label{eq3}
\end{eqnarray}
where interactions \eqref{Int_i}--\eqref{Int_j} are given in Table \ref{Def_int} in terms of the density parameters.

\begin{table}[h!]
\begin{center}
\begin{tabular}{|l|}\hline
$\Gamma_{1T}=\frac{3}{2}\alpha H^2\left(1+\Omega_r-3\Omega_x\right)$\\\hline
$\Gamma_{1d}=\frac{3}{2}\alpha H^2\left(1+\Omega_r-3\Omega_x\right)\left(\Omega_c+\Omega_x\right)$\\\hline
$\Gamma_{1c}=\frac{3}{2}\alpha 
H^2\left(1+\Omega_r-3\Omega_x\right)\Omega_c$\\\hline
$\Gamma_{1x}=\frac{3}{2}\alpha 
H^2\left(1+\Omega_r-3\Omega_x\right)\Omega_x$\\\hline
$\Gamma_{2T}=-\frac{3}{2}\alpha H^2\left(1+\Omega_r-3\Omega_x\right)\left(3+\Omega_r-3\Omega_x\right)$\\\hline
$\Gamma_{2d}=-\frac{9}{2}\alpha H^2(1+\Omega_r-3\Omega_x)\Omega_c$\\\hline
$\Gamma_{2c}=-9\alpha H^2\frac{\Omega_c(1+\Omega_r-3\Omega_x)}{2+3\alpha(1+\Omega_r-3\Omega_x)}$
\\\hline
\end{tabular}
\caption{\label{Def_int} Revised sign-changeable interactions}
\end{center}
\end{table}

The stability of the relevant cosmological epochs is found by calculating the eigenvalues of the linearized system at the critical points. For each cosmological interaction we follow the following scheme. First we consider the set of equations,
\begin{eqnarray}
\Omega_i'=f_i(\Omega_l),\label{DS}
\end{eqnarray}
where $f_i$ is a function of the density parameters $\Omega_r$, $\Omega_c$ and $\Omega_x$ and $i=r,c,x$. From this we find the critical points $\Omega_l^*$ of the system \eqref{DS} by calculating 
\begin{equation*}
f_i(\Omega_l^*)=0.
\end{equation*}
Then we linearized the set of equations \eqref{DS} around the critical points,
\begin{eqnarray*}
\delta \Omega_i'=J_i^l(\Omega_j^*)\Omega_l,
\end{eqnarray*}
where $J_i^l=\frac{\partial f_i}{\partial \Omega_l}$ is the {jacobian matrix.} By analysing the jacobian matrix we can find regions in the parameter space where a specific behavior is presented. In particular, we have stable, saddle or unstable points when the real part of the eigenvalues are, all negatives, a mixture of positives and negatives or all positives, respectively.

In the following, we describe in detail each of the models in Table \ref{Def_int}. First we present models $\Gamma_{1T}$, $\Gamma_{1x}$ and $\Gamma_{2T}$ which can be reduced to two-dimensional systems. Subsequently we describe models $\Gamma_{1d}$, $\Gamma_{1c}$, $\Gamma_{2d}$ and $\Gamma_{2c}$, which are analysed in terms of three-dimensional systems.
We describe each of the critical points with the focus on radiation, matter and dark energy dominance, corresponding to an unstable, saddle and stable points, respectively.

\subsection{Case A: $\Gamma_{1T}$}\label{G1T}
For the model $\Gamma_{1T}$ the set of Eqs.\eqref{eq1}-\eqref{eq3} becomes,
\begin{eqnarray}
\Omega_r'&=&\Omega_r(-1+\Omega_r-3\Omega_x),\label{eqA1}\\
\Omega_c'&=&-\frac{3\alpha}{2} \left(1+\Omega_r-3\Omega_x\right)+\Omega_c(\Omega_r-3\Omega_x),\label{eqA2}\\
\Omega_x'&=&\frac{3\alpha}{2}\left(1+\Omega_r-3\Omega_x\right)+\Omega_x(3+\Omega_r-3\Omega_x).\label{eqA3}
\end{eqnarray}
We notice that the dynamical system analysis can be simplified given that Eqs.\eqref{eqA1} and \eqref{eqA3} become a closed system, then we can perform a 2-dimensional analysis. By considering $\Omega_r'=0$ and $\Omega_x'=0$ simultaneously we find the critical points of the system. In order to obtain the stability of the critical points we linearized the system around the critical points and we analyzed the corresponding eigenvalues.

For the model $\Gamma_{1T}$ we find the following three critical points $A_i=\{\Omega_{r},\Omega_{x}\}$,
\begin{eqnarray*}
A_1&=&\{1-\frac{9}{4}\alpha,-\frac{3}{4}\alpha\},\\
A_2&=&\left\{0,\frac{1}{4}\left(2-3\alpha-\sqrt{4-\alpha\left(4-9\alpha\right)}\right)\right\},\\
A_3&=&\left\{0,\frac{1}{4}\left(2-3\alpha+\sqrt{4-\alpha\left(4-9\alpha\right)}\right)\right\},
\end{eqnarray*}
where the matter contribution to the critical points $\Omega_m=\Omega_b+\Omega_c$ is determined by the constraint 
\begin{equation}
\Omega_r+\Omega_m+\Omega_x=1.
\label{constraint}
\end{equation}
In the physical analysis of these points we have considered as existence conditions $\Omega_r\ge0$ and $\Omega_m\ge0$ at the critical points. On the other hand, notice that the effective state parameter $\omega_{\rm eff}=\frac{p_T}{\rho_T}$ ($p_T$ is the total pressure) is related to the deceleration parameter by $q=\frac{1}{2}(1+3\omega_{\rm eff})$, and in terms of the density parameters $q$ is given by
\begin{equation}
q=\frac{1}{2}(1+\Omega_r-3\Omega_x). \label{qa}
\end{equation}

Table \ref{T2} shows the existence and stability conditions on the eigenvalues $A_1-A_3$, as well as the $\omega_{\rm eff}$ value. The critical point $A_1$ in Table \ref{T2} describes a radiation dominated epoch where radiation, matter and dark energy coexist for $0<\alpha<\frac{4}{9}$.  The matter contribution to this critical point is given by $\Omega_m=3\alpha$ and in this case $\Omega_x$ is always negative given the existence conditions. The signs of the eigenvalues of the linearized system indicate that $A_1$ is an unstable critical point for $\alpha<\frac{4}{9}$.

\begin{table*}[ht!]
\begin{center}\begin{minipage}{\textwidth}
\centering
\begin{tabular}{|c|c|c|c|}\hline
Critical Points & $\omega_{\rm eff}$ & Existence & Stability Conditions\\\hline

\multirow{2}{*}{$A_1$} & \multirow{2}{*}{$\frac{1}{3}$} &
\multirow{2}{*}{$0<\alpha<\frac{4}{9} $} & 
{\bf unstable} for $ \alpha<\frac{4}{9}$\\
& && saddle for $\alpha>\frac{4}{9}$\\ \cline{1-4}
\multirow{2}{*}{$A_2$} & 
\multirow{2}{*}{$-\frac{1}{2}+\frac{1}{2}\sqrt{1-\frac{\alpha\left(4-9\alpha\right)}{4}}+\frac{3\alpha}{4}$} &
\multirow{2}{*}{$\alpha \in \mathbb{R}$}& 
unstable for $\alpha>\frac{4}{9}$ \\
& && {\bf saddle} for $\alpha<\frac{4}{9}$\\\cline{1-4}
$A_3$  & $-\frac{1}{2}-\frac{1}{2}\sqrt{1-\frac{\alpha\left(4-9\alpha\right)}{4}}+\frac{3\alpha}{4}$ & $\alpha >0$ & {\bf stable} \\ \hline
\end{tabular}
\end{minipage}
\end{center}
\caption{\label{T2} Description of the critical points for scenario $\Gamma_{1T}$.}
\end{table*}

The critical point $A_2$ has $\omega_{\rm eff}>-\frac{1}{3}$ irrespective of the value of $\alpha$. In the limit $\alpha\rightarrow0$, this phase can be understood as a phase of matter domination, given that $\omega_{\rm eff}\rightarrow0$ in this limit. In this phase  matter and dark energy coexists. At this critical point the matter component is $\Omega_m=\frac{1}{4}\left(2+3\alpha+\sqrt{4-\alpha(4-9\alpha)}\right)$, which is positive irrespective of the $\alpha$ value. On the other hand, we have $\Omega_x<0$ for $\alpha>0$ and $\Omega_x>0$ for $\alpha<0$. $A_2$ is a saddle critical point for $\alpha<\frac{4}{9}$ or an unstable critical point for $\alpha>\frac{4}{9}$.

The critical point $A_3$ dynamically corresponds to a phase of dark energy domination ($\omega_{\rm eff}< -\frac{1}{3}$ irrespective of the $\alpha$ value), where matter and dark energy coexist. {At this point $\Omega_m=\frac{1}{4}\left(2+3\alpha-\sqrt{4-\alpha(4-9\alpha)}\right)$, which is positive for $\alpha>0$}. Under this existence condition $\Omega_x$ is always positive. $A_3$ is a stable critical point for $\alpha>0$.

\begin{figure}[ht!]
\begin{center}
\centering
\includegraphics[width=0.45\textwidth]{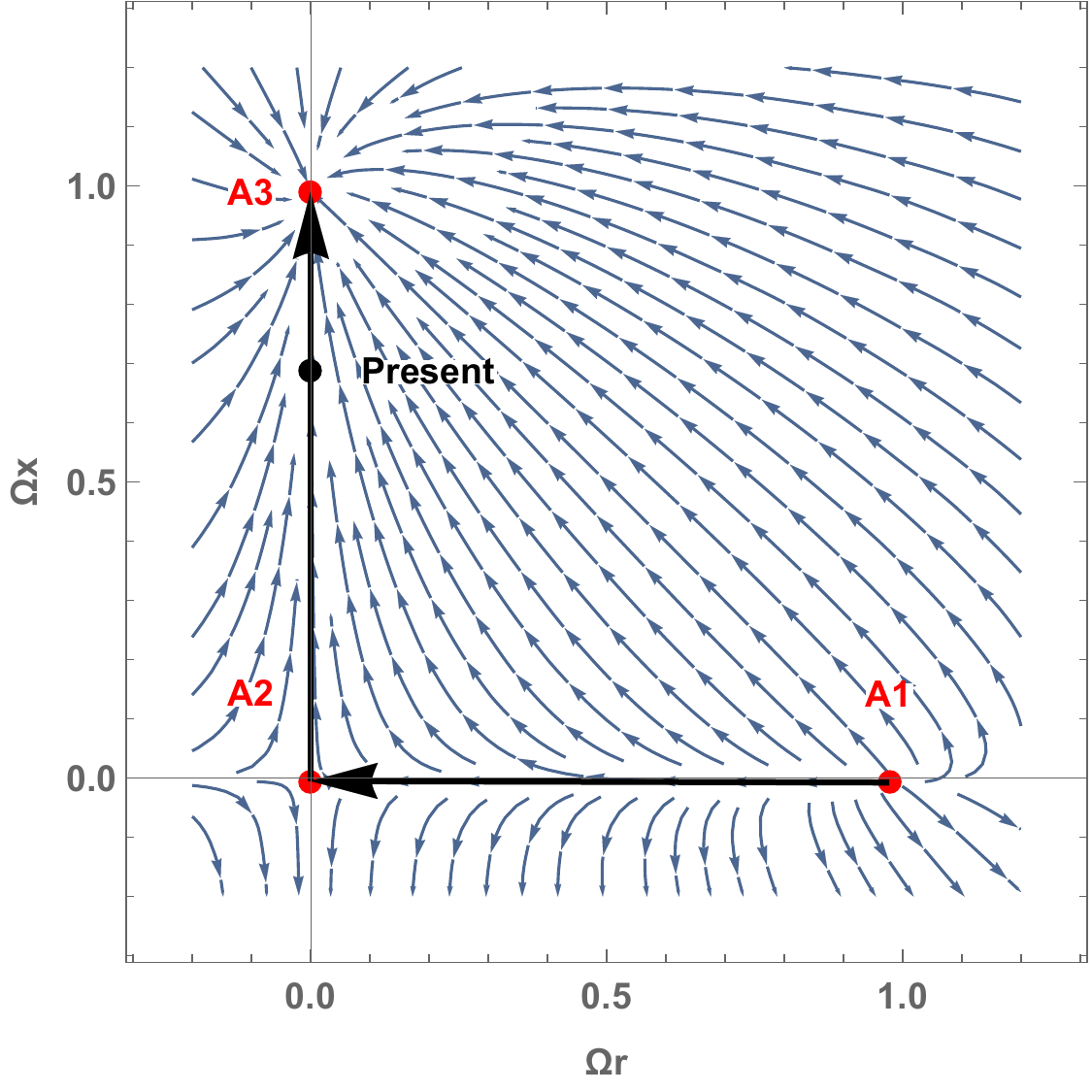}
\caption{\label{Fig1}Phase plot for model $\Gamma_{1T}$ with $\alpha=0.01$.} 
\end{center}
\end{figure}
Figure \ref{Fig1} shows the 2-dimensional phase space $\Omega_x-\Omega_r$. The three critical points in Table \ref{T2} are shown as red dots, where we can clearly notice the unstable, saddle and stable points, these 3 points can be find in model $\Gamma_{1T}$ for {$0\le\alpha<\frac{4}{9}$}, nevertheless, in this range $\Omega_x$ is negative at points $A_1$ and $A_2$. We also show a possible trajectory  going through the three critical points, starting at radiation domination and ending at dark energy domination. In the figure, the ``Present'' point is indicated considering $\{\Omega_m,\Omega_x\}=\{0.3111,0.6889\}$ \cite{Planck:2018vyg}.

Finally notice that the critical points $A_1-A_3$ in table \ref{T2} and their existence and stability conditions reduce to the ones in the $\Lambda$CDM scenario in the limit $\alpha\rightarrow0$.

\subsection{Case B: $\Gamma_{1x}$}
For the model $\Gamma_{1x}$ we can  reduce the set of Eqs.\eqref{eq1}-\eqref{eq3} to a 2-dimensional set including Eq.\eqref{eq1} and the following equation,
\begin{eqnarray*}
\Omega_x'=\Omega_x\left(\frac{3\alpha}{2}\left(1+\Omega_r-3\Omega_x\right)+3+\Omega_r-3\Omega_x\right).\label{eqB1}
\end{eqnarray*}
For this model we find three critical points $B_i=\{\Omega_r,\Omega_x\}$,
\begin{eqnarray*}
B_1=\{1,0\},\quad
B_2=\left\{0,0\right\},\quad
B_3=\left\{0,\frac{2+\alpha}{2+3\alpha}\right\},
\end{eqnarray*}
where the matter contribution to the critical points is determined by the constraint \eqref{constraint}.

The existence and stability conditions of the critical points $B_1-B_3$ are shown in Table \ref{T3}, as well as the $\omega_{\rm eff}$ parameter.

\begin{table*}\begin{center}\begin{minipage}{\textwidth}
\centering
\begin{tabular}{|c|c|c|c|}\hline
Critical Points & $\omega_{\rm eff}$ & Existence & Stability Conditions\\\hline
\multirow{2}{*}{$B_1$} & 
\multirow{2}{*}{$\frac{1}{3}$} &\multirow{2}{*}{$\alpha \in \mathbb{R}$}& {\bf unstable} $\alpha>-\frac{4}{3}$  \\ 
& && saddle $\alpha <-\frac{4}{3}$\\\cline{1-4}
\multirow{2}{*}{$B_2$}& \multirow{2}{*}{{$0$}}&\multirow{2}{*}{$\alpha\in \mathbb{R}$}& {stable} for $\alpha<-2$\\
& && {\bf saddle} for $\alpha>-2$ \\\cline{1-4}
\multirow{2}{*}{$B_3$} & \multirow{2}{*}{$-\frac{2+\alpha}{2+3\alpha}$} &\multirow{2}{*}{$\alpha>0$ or $\alpha<-\frac{2}{3}$}& {\bf stable} for $\alpha>-\frac{2}{3}$  or $ -2<\alpha<-\frac{4}{3}$ \\ 
& && {saddle} for $\alpha<-2$ or $ -\frac{4}{3}<\alpha<-\frac{2}{3}$ \\\hline
\end{tabular}
\end{minipage}
\caption{\label{T3} Description of the critical points for scenario $\Gamma_{1x}$.}
\end{center}
\end{table*}

The critical points $B_1$ and $B_2$ in Table \ref{T3} corresponds to a phase of radiation and matter domination, respectively. There are no conditions on the existence of these critical points and the stability analysis indicate that $B_1$ is unstable for $\alpha>-\frac{4}{3}$ and saddle for $\alpha<-\frac{4}{3}$, meanwhile, $B_2$ is saddle for $\alpha>-2$ and stable for $\alpha<-2$.

At the critical point $B_3$ matter and dark energy coexist. At this point $\Omega_m=\frac{2\alpha}{2+3\alpha}$ which is positive for $\alpha<-\frac{2}{3}$ or $\alpha>0$. On the other hand, for $\alpha>0$, this point corresponds to dark energy domination with $\Omega_x>0$ as a stable critical point. For $\alpha<-\frac{2}{3}$ we get $\omega_{\rm eff}<-\frac{1}{3}$, $\Omega_m>\Omega_x$ and $\Omega_x>0$ for $\alpha<-2$. The corresponding stability conditions in this case are given in Table \ref{T3}.

\begin{figure}[ht!]
\begin{center}
\centering
 \includegraphics[width=0.45\textwidth]{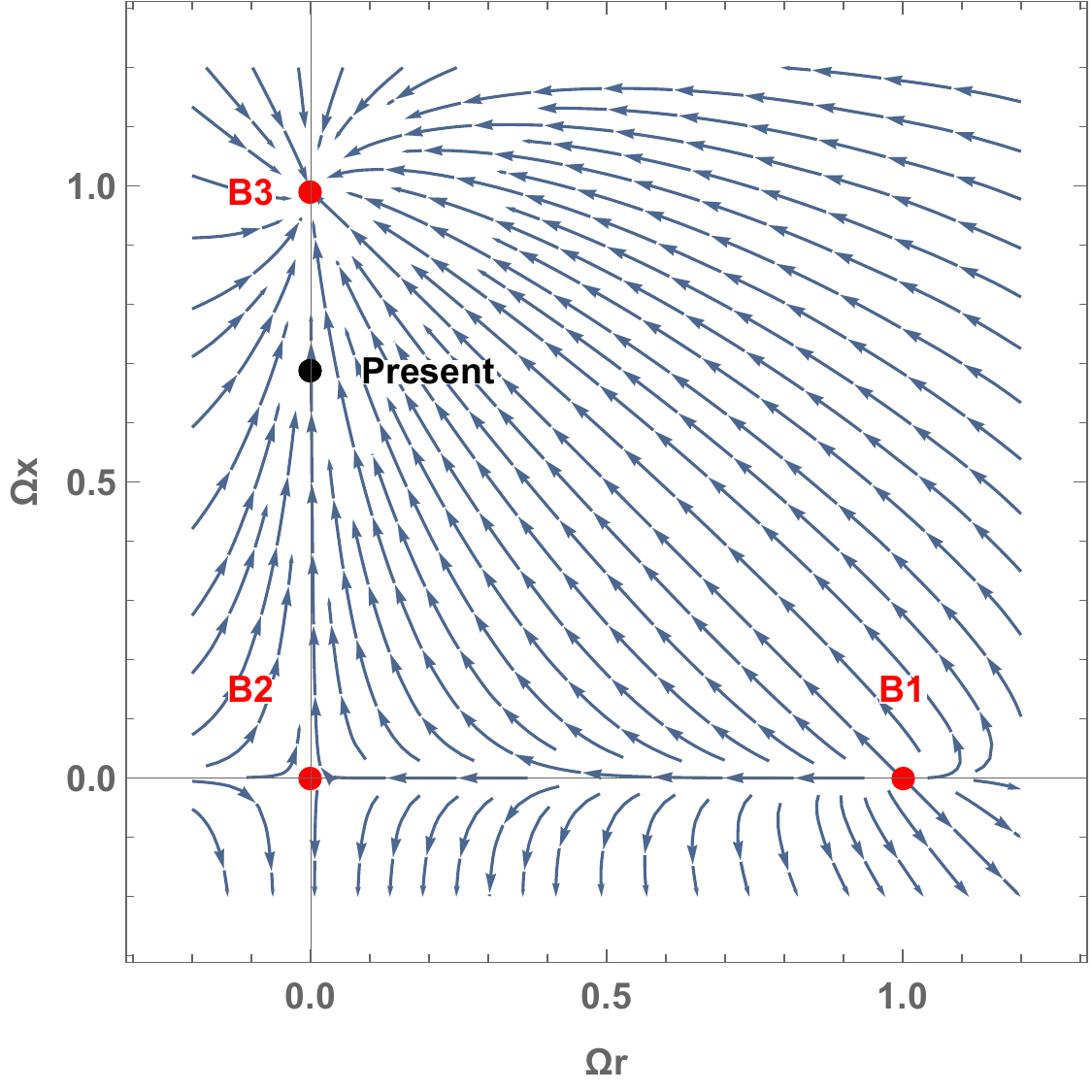}
    \caption{Phase plot for model $\Gamma_{1x}$ with $\alpha=0.01$.\label{Fig2}} 
\end{center}
\end{figure}

The figure \ref{Fig2} shows the 2-dimensional phase space $\Omega_x-\Omega_r$ for the $\Gamma_{1x}$ scenario, where the three critical points in Table \ref{T3} are shown as red dots and we can clearly notice the unstable, saddle and stable points. These points can be find in model $\Gamma_{1x}$ for $\alpha\ge0$, where $\Omega_x$ is positive at the critical point $B_3$. In the figure the black dot labeled as ``Present'' considers the values in Ref.\cite{Planck:2018vyg}.

As well as model $\Gamma_{1T}$, the critical points $B_1-B_3$ reduce to the $\Lambda$CDM critical points in the limit $\alpha \rightarrow0$.

\subsection{Case C: $\Gamma_{2T}$}
For the model $\Gamma_{2T}$ we can  reduce the set of Eqs.\eqref{eq1}-\eqref{eq3} to a 2-dimensional set including Eq.\eqref{eq1} and the following equation,
\begin{eqnarray*}
\Omega_x'=\left(3+\Omega_r-3\Omega_x\right)\left(\Omega_x-\frac{3\alpha}{2}\left(1+\Omega_r-3\Omega_x\right)\right).\label{eqC1}
\end{eqnarray*} 
For this model we find three critical points $C_i=\{\Omega_r,\Omega_x\}$,
\begin{eqnarray*}
&&C_1=\{1+9\alpha,3\alpha\},\,\quad
C_2=\left\{0,\frac{3\alpha}{2+9\alpha}\right\},\,\quad
C_3=\{0,1\},
\end{eqnarray*}
where the matter contribution to the critical points is determined by the constraint \eqref{constraint}.

The existence and stability conditions of the critical points are shown in Table \ref{T4}, as well as the $\omega_{\rm eff}$ parameter.

At the critical point $C_1$ coexist radiation, matter and dark energy in a radiation dominated era ($\omega_{\rm eff}=\frac{1}{3}$). For this critical point we have $\Omega_m=-12\alpha$ which is positive for $\alpha<0$ but necessarily in this case $\Omega_x<0$.
At the critical point $C_2$ coexist matter and dark energy, $\omega_{\rm eff}\ge-\frac{1}{3}$ for $\alpha>-\frac{2}{9}$. At this point we have $\Omega_m=\frac{2+6\alpha}{2+9\alpha}$ which is positive for $\alpha<-\frac{1}{3}$ or $\alpha>-\frac{2}{9}$. In the range $\alpha<-\frac{2}{9}$ or $\alpha>0$, the parameter $\Omega_x$ is also positive. This point corresponds to matter domination in the limit $\alpha\rightarrow0$.
\begin{figure}[ht!]
\begin{center}
\centering
 \includegraphics[width=0.45\textwidth]{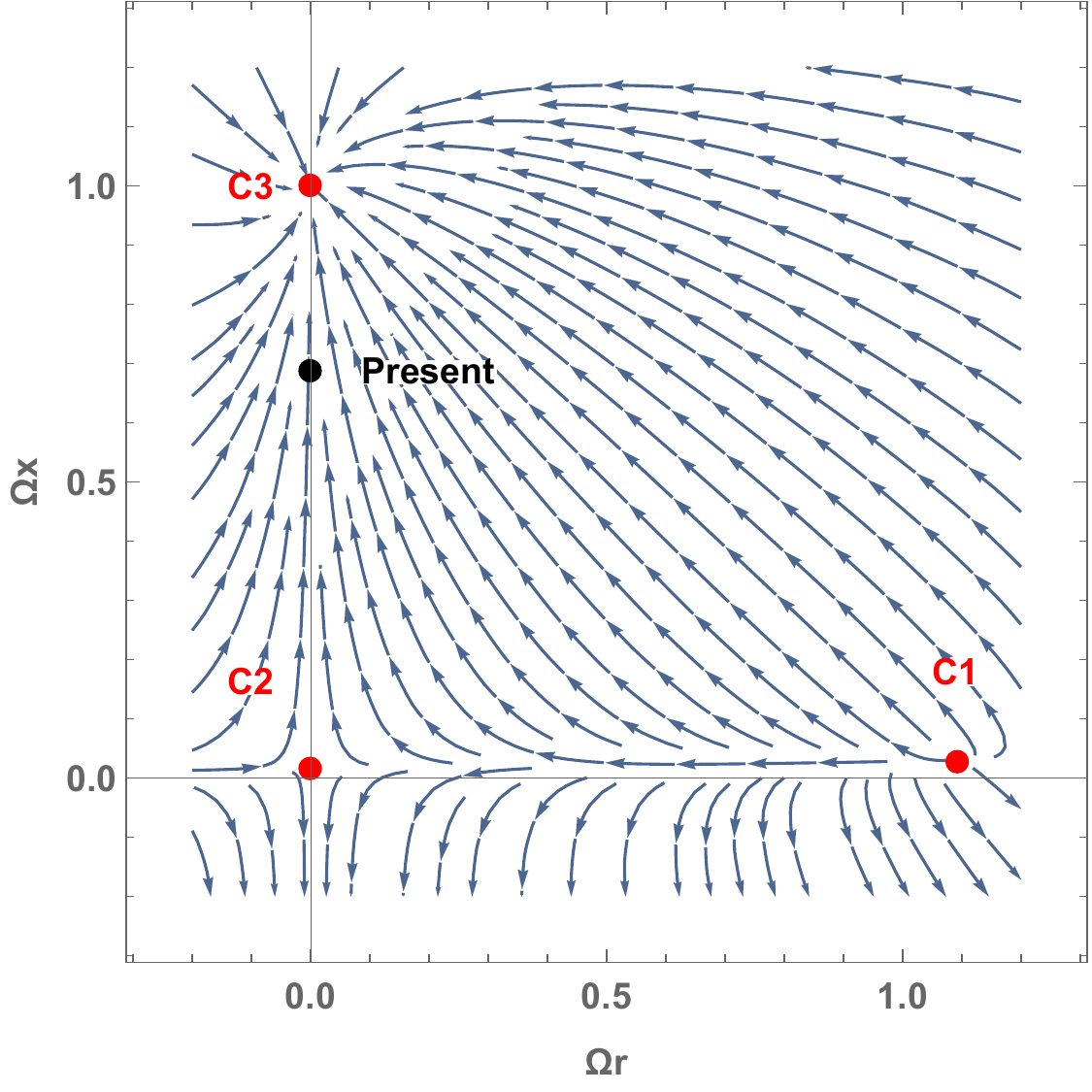}
    \caption{Phase plot for model $\Gamma_{2T}$ with $\alpha=-0.01$.\label{Fig3}} 
\end{center}
\end{figure}
The point $C_3$ corresponds to a de-Sitter attractor for $\alpha>-\frac{1}{3}$, where only the dark energy term contributes.

The figure \ref{Fig3} shows the 2-dimensional phase space $\Omega_x-\Omega_r$ for the $\Gamma_{2T}$ scenario. The three critical points in Table \ref{T4} are shown as red dots, we can clearly notice the unstable, saddle and stable points, which can be find in the model $\Gamma_{2T}$ for $-\frac{1}{9}<\alpha\le0$. However, in this range, $\Omega_x$ is negative at the critical points $C_1$ and $C_2$. In the figure the ``Present'' dot considers the values in Ref.\cite{Planck:2018vyg}.

\begin{table*}\begin{minipage}{\textwidth}
\centering
\begin{tabular}{|c|c|c|c|}\hline
Critical Points & $\omega_{\rm eff}$ & Existence & Stability Conditions\\\hline
\multirow{2}{*}{$C_1$} & \multirow{2}{*}{$\frac{1}{3}$}&\multirow{2}{*}{$-\frac{1}{9}<\alpha<0$}& {\bf unstable} for $\alpha>-\frac{1}{9}$ \\ 
& && saddle for $ \alpha<-\frac{1}{9}$\\\cline{1-4}
\multirow{3}{*}{$C_2$} & \multirow{3}{*}{$-\frac{3\alpha}{2+9\alpha }$} &
\multirow{3}{*}{{$\alpha>-\frac{2}{9}$\; or\; $\alpha<-\frac{1}{3}$} }& 
unstable for $-\frac{2}{9}<\alpha<-\frac{1}{9}$\\
& && stable for $\alpha<-\frac{1}{3}$\\
& && {\bf saddle} for $-\frac{1}{3}<\alpha<-\frac{2}{9}$ or $\alpha>-\frac{1}{9}$ \\\cline{1-4}
\multirow{2}{*}{$C_3$} & \multirow{2}{*}{{-1}} & \multirow{2}{*}{$\alpha\in\mathbb{R}$}&{\bf stable} for $\alpha>-\frac{1}{3}$  \\ 
& && saddle for $\alpha<-\frac{1}{3}$\\\hline

\end{tabular}
\end{minipage}
\caption{\label{T4} Description of the critical points for scenario $\Gamma_{2T}$.}
\end{table*}

The critical points $C_1-C_3$ corresponding to model $\Gamma_{2T}$ in Table \ref{T4} reduce to the $\Lambda$CDM critical points in the limit $\alpha\rightarrow0$.

\subsection{Case D: $\Gamma_{1d}$}
For the model $\Gamma_{1d}$ the set of Eqs.\eqref{eq1}-\eqref{eq3} reduces to,
\begin{eqnarray*}
\Omega_r'&=&\Omega_r(-1+\Omega_r-3\Omega_x),\label{eqD1}\\
\Omega_c'&=&\Omega_c(\Omega_r-3\Omega_x)-\frac{3\alpha}{2}\left(1+\Omega_r-3\Omega_x\right)\left(\Omega_c+\Omega_x\right),\label{eqD2}\\
\Omega_x'&=&\Omega_x(3+\Omega_r-3\Omega_x)+\frac{3\alpha}{2}\left(1+\Omega_r-3\Omega_x\right)\left(\Omega_c+\Omega_x\right).\label{eqD3}
\end{eqnarray*}
For this model we find the following critical points $D_i=\{\Omega_r,\Omega_b,\Omega_x\}$,
\begin{eqnarray*}
D_1=\{1,0,0\},\quad
D_2=\{0,1,0\},\\
D_3=\left\{0,0,\frac{1}{4}\left(2-3\alpha-\sqrt{4-\alpha\left(4-9\alpha\right)}\right)\right\},\\
D_4=\left\{0,0,\frac{1}{4}\left(2-3\alpha+\sqrt{4-\alpha\left(4-9\alpha\right)}\right)\right\},
\end{eqnarray*}
where the $\Omega_c$ contribution to the critical points is determined by the constraint \eqref{const0}.

In the physical analysis of these points we have considered as existence conditions $\Omega_r\ge0$, $\Omega_b\ge0$ and $\Omega_c\ge0$ at the critical points.

 \begin{table*}\begin{minipage}{\textwidth}
\centering
\begin{tabular}{|c|c|c|c|}\hline
Critical Points & $\omega_{\rm eff}$ & Existence & Stability Conditions\\\hline
 \multirow{2}{*}{$D_1$} & \multirow{2}{*}{$\frac{1}{3}$} &\multirow{2}{*}{$\alpha\in \mathbb{R}$}& {\bf unstable} for $-\frac{1}{4}\le \alpha<\frac{4}{9}$\\
 & && saddle for $\alpha>\frac{4}{9}$ \\ \cline{1-4}%
 \multirow{2}{*}{$D_2$} & \multirow{2}{*}{$0$} &\multirow{2}{*}{$\alpha\in \mathbb{R}$} & stable for $\alpha=\frac{8}{9}$\\ 
 & && saddle for $-\frac{1}{2}\le\alpha<\frac{8}{9}$ or $\alpha>\frac{8}{9}$\\\cline{1-4}%
 \multirow{2}{*}{$D_3$} & \multirow{2}{*}{$-\frac{1}{2}+\frac{1}{2}\sqrt{1-\frac{\alpha\left(4-9\alpha\right)}{4}}+\frac{3\alpha}{4}$} &\multirow{2}{*}{$\alpha\in \mathbb{R}$}& unstable for $\alpha>\frac{4}{9}$ \\
 & && {\bf saddle} for $\alpha<\frac{4}{9}$\\\cline{1-4} %
 $D_4$  & $-\frac{1}{2}-\frac{1}{2}\sqrt{1-\frac{\alpha\left(4-9\alpha\right)}{4}}+\frac{3\alpha}{4}$ & $\alpha>0$ & {\bf stable} \\\hline
 \end{tabular}
 \end{minipage}
 \caption{\label{T5} Description of the critical points for scenario $\Gamma_{1d}$.}
 \end{table*}
 The critical points $D_1$ and $D_2$ in Table \ref{T5} describe a radiation and a baryon dominated epoch, respectively, irrespective of the $\alpha$ value. The signs of the eigenvalues of the linearized system indicate that $D_1$ is an unstable critical point in the range $-\frac{1}{4}\le\alpha<\frac{4}{9}$. On the other hand, the critical point $D_2$ corresponds to a saddle point in the above range but it represents a non-physical point because there is not a baryon dominated epoch in the actual evolution of our universe.

We notice that scenarios $\Gamma_{1T}$ and $\Gamma_{1d}$ share the same critical points for the dark sector, $D_3$ and $D_4$, with the same existence and stability conditions. 
 
\begin{figure}[ht!]
\centering
\includegraphics[width=0.45\textwidth]{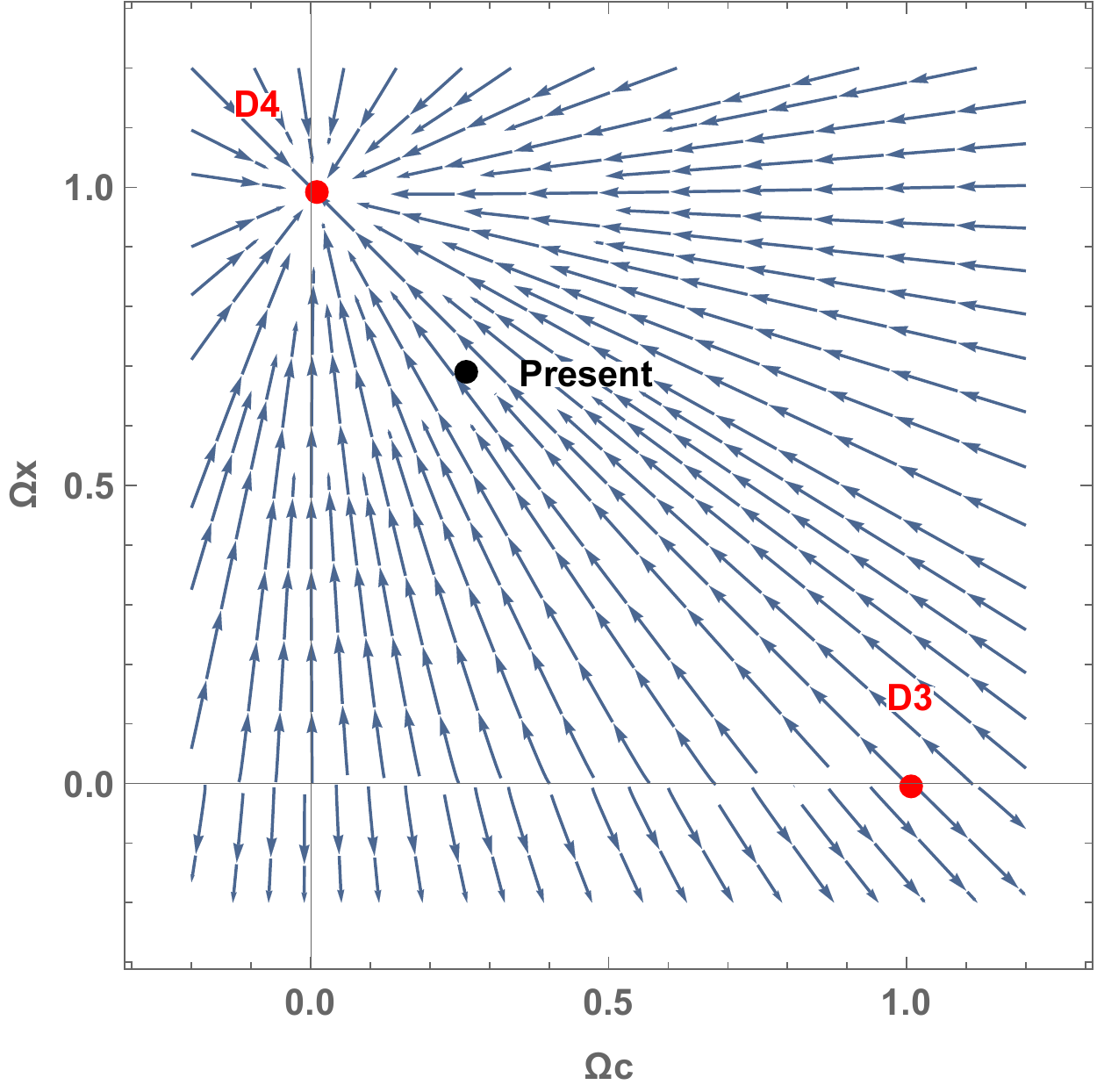} 
\caption{Projected phase plot for model $\Gamma_{1d}$ with $\alpha=0.01$  \label{Fig4}}
\end{figure}

The critical point $D_3$ can be understood as dark matter domination in the limit $\alpha\rightarrow 0$, this point (and its analysis) is equivalent to the point $A_2$ for the scenario $\Gamma_{1T}$. The critical point $D_4$ corresponds to dark energy domination for $\alpha>0$. This point and its analysis is equivalent to the points $A_3$ in scenario $\Gamma_{1T}$. The existence conditions in Table \ref{T5} indicate that these critical points (and the radiation one) can be find for $0\le\alpha<\frac{4}{9}$, but in this range necessarily $\Omega_x$ becomes negative at the critical point $D_3$. In Fig.\ref{Fig4} we show a projected 2-dimensional phase space $\Omega_x-\Omega_c$ where we have fixed $\Omega_r=0$. In this figure we show two critical points (physical) corresponding to the saddle point $D_3$ and the dark energy attractor $D_4$.

\subsection{Case E: $\Gamma_{1c}$}
For the model $\Gamma_{1c}$ the set of Eqs.\eqref{eq1}-\eqref{eq3} reduces to,
\begin{eqnarray*}
\Omega_r'&=&\Omega_r(-1+\Omega_r-3\Omega_x),\label{eqc1}\\
\Omega_c'&=&\Omega_c(\Omega_r-3\Omega_x)-\frac{3}{2}\alpha 
\left(1+\Omega_r-3\Omega_x\right)\Omega_c,\label{eqc2}\\
\Omega_x'&=&\Omega_x(3+\Omega_r-3\Omega_x)+\frac{3}{2}\alpha \left(1+\Omega_r-3\Omega_x\right)\Omega_c.\label{eqc3}
\end{eqnarray*}
For this model we find the following critical points $E_i=\{\Omega_r,\Omega_b,\Omega_x\}$,
\begin{eqnarray*}
E_1=\{1,0,0\},\quad
E_2=\{0,1,0\},\quad
E_3=\left\{0,0,-\frac{\alpha}{2-3\alpha}\right\},\quad
E_4=\left\{0,0,1\right\},
\end{eqnarray*}
where the contribution $\Omega_c$ to the critical points is determined by the constraint \eqref{const0} to be zero for each point but $E_3$, where $\Omega_c=\frac{2(1-\alpha)}{2-3\alpha}$.

\begin{table*}\begin{minipage}{\textwidth}
\centering
\begin{tabular}{|c|c|c|c|}\hline
Critical Points & $\omega_{\rm eff}$ & Existence & Stability Conditions\\\hline
\multirow{2}{*}{$E_1$} & \multirow{2}{*}{$\frac{1}{3}$} &\multirow{2}{*}{$\alpha \in \mathbb{R}$}& {\bf unstable} for $\alpha<\frac{1}{3}$ \\
& && saddle for $\alpha >\frac{1}{3}$\\\cline{1-4}
\multirow{2}{*}{$E_2$} & \multirow{2}{*}{$0$}& \multirow{2}{*}{$\alpha \in \mathbb{R}$}& stable for $\alpha=\frac{2}{3}$ \\ 
& && saddle for $\alpha\neq\frac{2}{3}$\\\cline{1-4}
\multirow{3}{*}{$E_3$} & \multirow{3}{*}{$\frac{\alpha}{2-3\alpha}$} & \multirow{3}{*}{$\alpha<\frac{2}{3}$ \;or\;$\alpha>1$}&unstable for $\frac{1}{3}<\alpha<\frac{2}{3}$\\
& && stable for $\alpha>1$\\
& && {\bf saddle} for $\alpha<\frac{1}{3}$ or $\frac{2}{3}<\alpha<1$\\\cline{1-4}
\multirow{2}{*}{$E_4$} &  \multirow{2}{*}{-1}&\multirow{2}{*}{$\alpha \in \mathbb{R}$}& {\bf stable} for $\alpha<1$\\
 & && saddle for $\alpha>1$
\\\hline
\end{tabular}
\end{minipage}
\caption{\label{T6}Description of the critical points for scenario $\Gamma_{1c}$.}
\end{table*}
 
\begin{figure}[ht]
\centering
\includegraphics[width=0.4\textwidth]{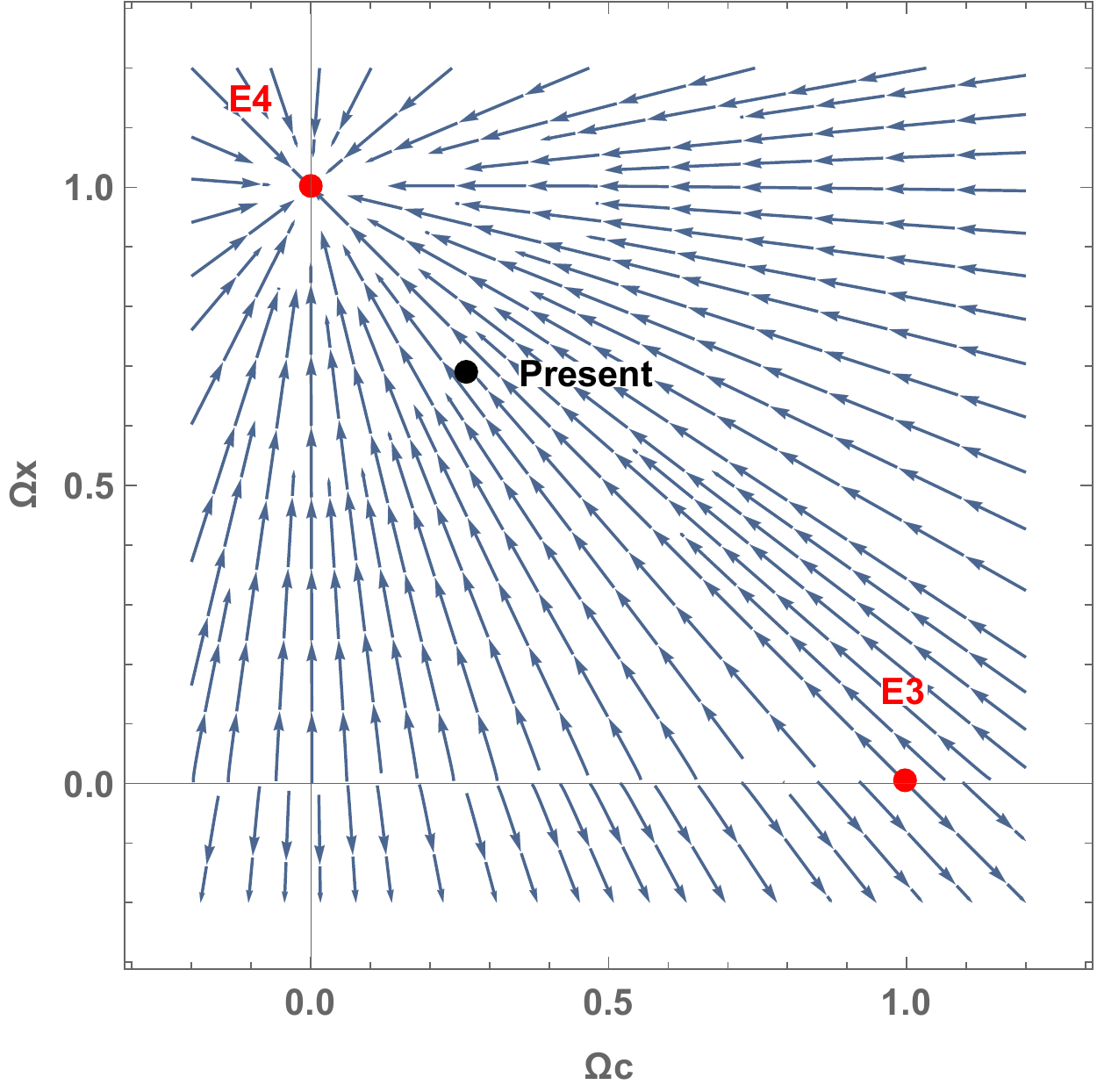}
\caption{Projected phase plot for model $\Gamma_{1c}$ with $\alpha=-0.01$ \label{Fig5}}
\end{figure}

The critical points $E_1$ and $E_2$ in Table \ref{T6} describe a radiation and a baryon dominated epoch, respectively. The signs of the eigenvalues of the linearized system indicate that $E_1$ is an unstable critical point for $\alpha<\frac{1}{3}$. On the other hand, the non-physical point $E_2$ (baryon domination) corresponds to a saddle point in the above range. The point $E_3$ represent a phase where the dark fluids coexist with $\Omega_c>0$ for $\alpha<\frac{2}{3}$ or $\alpha>1$ and $\omega_{\rm eff}>-\frac{1}{3}$ for $\alpha<\frac{2}{3}$, this point corresponds to dark matter domination in the limit $\alpha\rightarrow0$. The point $E_4$ corresponds to a de-Sitter stage with the dominance of dark energy. This point is stable for $\alpha<1$.
 
From Table \ref{T6} we notice that the critical points corresponding to unstable, saddle and stable behavior can be find for $\alpha<\frac{1}{3}$. Considering this range,  $\Omega_x$ remains positive for $\alpha<0$ at the critical point $E_3$. In Fig.\ref{Fig5} we show the projected 2-dimensional phase space $\Omega_x - \Omega_c$ where we have fixed $\Omega_r=0$ and we have chosen $\alpha=-0.01$. In this figure we show two critical points (physical) corresponding to the saddle point $E_3$ and the attractor point $E_4$.

\subsection{Case F: $\Gamma_{2d}$}
For the model $\Gamma_{2d}$ the set of Eqs.\eqref{eq1}-\eqref{eq3} reduces to,
\begin{eqnarray*}
\Omega_r'&=&\Omega_r(-1+\Omega_r-3\Omega_x),\label{eq_2d1}\\
\Omega_c'&=&\frac{9}{2}\alpha(1+\Omega_r-3\Omega_x)\Omega_c+\Omega_c(\Omega_r-3\Omega_x),\label{eq_2d2}\\
\Omega_x'&=&-\frac{9}{2}\alpha(1+\Omega_r-3\Omega_x)\Omega_c+\Omega_x(3+\Omega_r-3\Omega_x).\label{eq_2d3}
\end{eqnarray*}

For this model we find the following critical points $F_i=\{\Omega_r,\Omega_b,\Omega_x\}$,
\begin{eqnarray*}
F_1=\{1,0,0\},\quad
F_2=\{0,1,0\},\quad
F_3=\left\{0,0,\frac{3\alpha}{2+9\alpha}\right\},\quad
F_4=\left\{0,0,1\right\},
\end{eqnarray*}
where the contribution $\Omega_c$ to the critical points is determined by the constraint \eqref{const0} to be zero at each point but $F_3$, where $\Omega_c=\frac{2+6\alpha}{2+9\alpha}$.

The critical points $F_1$ and $F_2$ in Table \ref{T7} describe a radiation and a baryon dominated era, respectively. The signs of the eigenvalues of the linearized system indicate that $F_1$ is an unstable critical point for $\alpha>-\frac{1}{9}$. On the other hand, the non-physical point $F_2$ (baryon domination) corresponds to a saddle point in the above range. The point $F_3$ represents a phase where the dark fluids coexist with $\Omega_c>0$ for $\alpha<-\frac{1}{3}$ or $\alpha>-\frac{2}{9}$ and $\omega_{\rm eff}>-\frac{1}{3}$ for $\alpha>-\frac{2}{9}$, this point corresponds to dark matter domination in the limit $\alpha\rightarrow0$. The point $F_3$ is saddle in the range $\alpha>-\frac{1}{9}$. The point $F_4$ corresponds to a de-Sitter stage with the dominance of dark energy and this point is stable for $\alpha>-\frac{1}{3}$.
 
\begin{figure}[ht]
\centering
\includegraphics[width=0.4\textwidth]{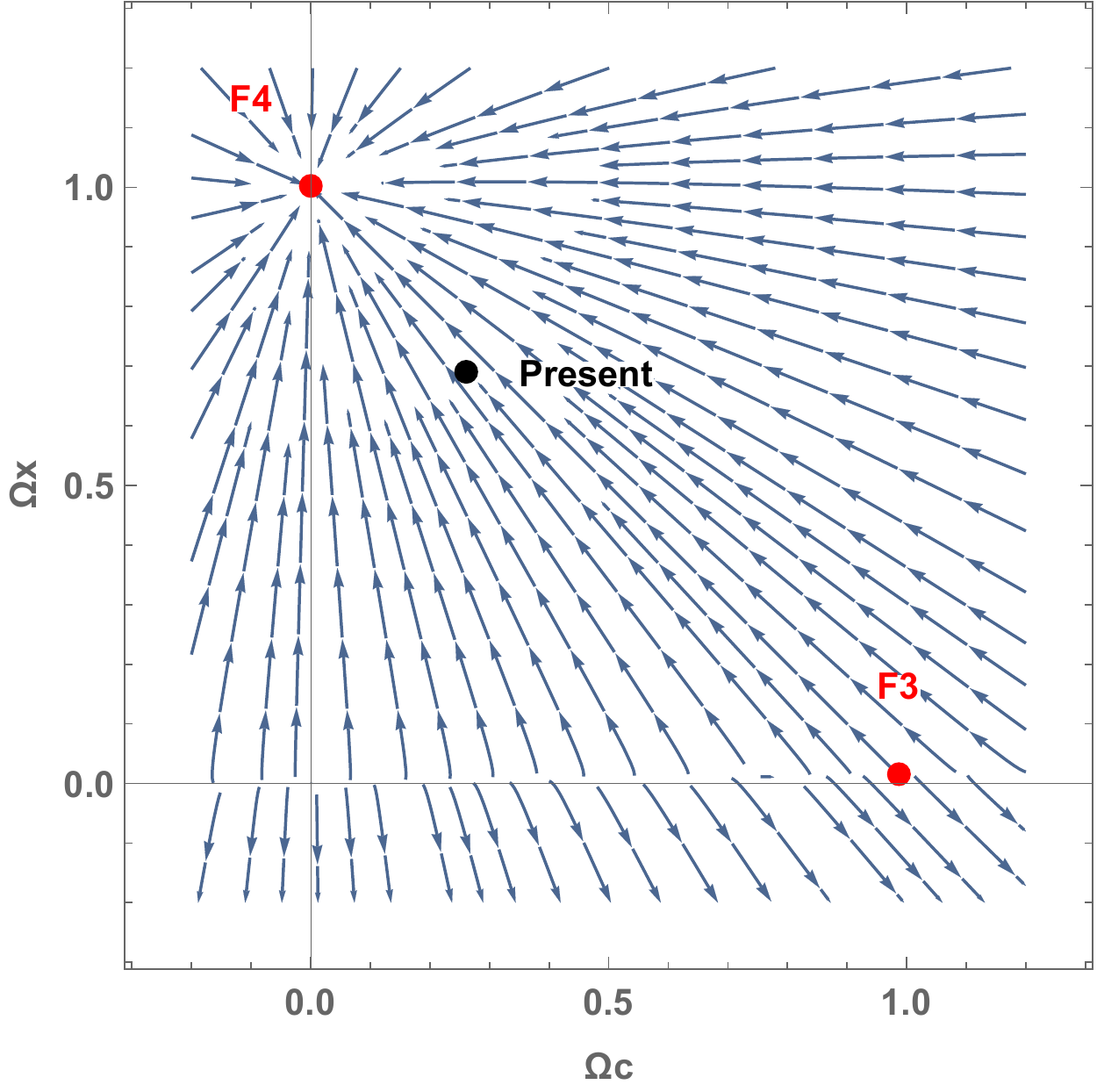} 
\caption{Projected phase plot for model $\Gamma_{2d}$ with $\alpha=0.01$ \label{Fig6}}
\end{figure}

From Table \ref{T7} we notice that the critical points corresponding to unstable, saddle and stable can be find for $\alpha>-\frac{1}{9}$. Notice that in this range,  $\Omega_x$ remains positive for $\alpha>0$ at the critical point $F_3$. In Fig.\ref{Fig6} we show the projected 2-dimensional phase space $\Omega_x - \Omega_c$ where we have fixed $\Omega_r=0$ and we have chosen $\alpha=0.01$. In this figure we show two critical points (physical) corresponding to the saddle point $F_3$ and the attractor point $F_4$.
Notice that the models $\Gamma_{2T}$ and $\Gamma_{2d}$ share the same critical points and existence/stability conditions in the points corresponding to the dark sector, $C_2-C_3$ and $F_3-F_4$. 

\begin{table*}\begin{minipage}{\textwidth}
\centering
\begin{tabular}{|c|c|c|c|}\hline
Critical Points & $\omega_{\rm eff}$ & Existence & Stability Conditions\\\hline
\multirow{2}{*}{$F_1$} & 
\multirow{2}{*}{$\frac{1}{3}$} &
\multirow{2}{*}{$\alpha\in\mathbb{R}$}&
{\bf unstable} for $\alpha>-\frac{1}{9}$\\
& && saddle for $\alpha<-\frac{1}{9}$ \\ \cline{1-4}
\multirow{2}{*}{$F_2$} & \multirow{2}{*}{$0$}&
\multirow{2}{*}{$\alpha\in\mathbb{R}$}& 
stable for $\alpha=-\frac{2}{9}$ \\
& && saddle for $ 2+9\alpha\neq0$\\\cline{1-4}
\multirow{3}{*}{$F_3$} & \multirow{3}{*}{$-\frac{3\alpha}{2+9\alpha}$} &
\multirow{3}{*}{$\alpha>-\frac{2}{9}$\; or\; $\alpha<-\frac{1}{3}$}& 
unstable for $-\frac{2}{9}<\alpha<-\frac{1}{9}$\\
& && stable for $\alpha<-\frac{1}{3}$\\
& && {\bf saddle} for $-\frac{1}{3}<\alpha<-\frac{2}{9}$ or $\alpha>-\frac{1}{9}$ \\\cline{1-4}
\multirow{2}{*}{$F_4$} & \multirow{2}{*}{{-1}} &\multirow{2}{*}{$\alpha\in\mathbb{R}$}& {\bf stable} for $\alpha>-\frac{1}{3}$  \\  & &&saddle for $\alpha<-\frac{1}{3}$\\\hline 
\end{tabular}
\end{minipage}
\caption{\label{T7}Description of the critical points for scenario $\Gamma_{2d}$.}
\end{table*}

\subsection{Case G: $\Gamma_{2c}$}
For the model $\Gamma_{2c}$ the set of Eqs.\eqref{eq1}-\eqref{eq3} reduces to,
\begin{eqnarray*}
\Omega_r'&=&\Omega_r(-1+\Omega_r-3\Omega_x),\label{eq_2c1}\\
\Omega_c'&=&\frac{9\alpha \Omega_c(1+\Omega_r-3\Omega_x)}{2+3\alpha(1+\Omega_r-3\Omega_x)}+\Omega_c(\Omega_r-3\Omega_x),\label{eq_2c2}\\
\Omega_x'&=&-\frac{9\alpha \Omega_c(1+\Omega_r-3\Omega_x)}{2+3\alpha(1+\Omega_r-3\Omega_x)}+\Omega_x(3+\Omega_r-3\Omega_x).\label{eq_2c3}
\end{eqnarray*}
For this model we find the following critical points $G_i=\{\Omega_r,\Omega_b,\Omega_x\}$,
\begin{eqnarray*}
G_1=\{1,0,0\},\quad
G_2=\{0,1,0\},\quad
G_5=\left\{0,0,1\right\},\\
G_3=\left\{0,0,\frac{1-\sqrt{1+3\alpha(4+3\alpha)}+6\alpha}{9\alpha}\right\},\\
G_4=\left\{0,0,\frac{1+\sqrt{1+3\alpha(4+3\alpha)}+6\alpha}{9\alpha}\right\},
\end{eqnarray*}
where given Eq. \eqref{const0}, $\Omega_c$ is zero at $G_1$ and $G_2$, $\Omega_c=\frac{-1+\sqrt{1+3\alpha(4+3\alpha)}+3\alpha}{9\alpha}$ at $G_3$, and  $\Omega_c=-\frac{1+\sqrt{1+3\alpha(4+3\alpha)}-3\alpha}{9\alpha}$ at $G_4$.

The critical points $G_1$ and $G_2$ in Table \ref{T8} describe a radiation and a baryon dominated epoch, respectively. The signs of the eigenvalues of the linearized system indicate that $G_1$ is an unstable critical point for $\alpha<-\frac{1}{3}$ or $\alpha>-\frac{1}{12}$. The non-physical point $G_2$ (baryon domination) corresponds to a saddle point in the above range. 

At the critical point $G_3$ the dark sector coexist in the ranges $\alpha<-\frac{\sqrt{3}+2}{3}$, $\frac{\sqrt{3}-2}{3}<\alpha<0$ or $\alpha>0$.  Inside these limits $G_3$ can be unstable for $\frac{\sqrt{3}-2}{3}<\alpha<-\frac{1}{12}$ or saddle for $\alpha<-\frac{\sqrt{3}+2}{3}$ or $\alpha>-\frac{1}{12}$. We notice that $\omega_{\rm eff}>-\frac{1}{3}$ for $\alpha>\frac{\sqrt{3}-2}{3}$ and this point reduces to dark matter domination in the limit $\alpha\rightarrow0$. Besides, $\Omega_x>0$ for $\alpha>0$ or $\alpha\le-\frac{\sqrt{3}+2}{3}$.
At the point $G_4$ coexist the dark components in the range $\frac{\sqrt{3}-2}{3}<\alpha<0$, where this point is saddle and $\alpha\le-\frac{\sqrt{3}+2}{3}$ where it is a stable critical point. In the range where $G_4$ is a saddle point we have $\omega_{\rm eff}>-\frac{1}{3}$, outside this range, in the allowed region, the effective fluid accelerate the universe's expansion.  The $\Omega_x$ contribution of this point is positive for $\alpha<-\frac{\sqrt{3}+2}{3}$. Notice that the point $G_4$ diverges in the limit $\alpha\rightarrow 0$. The point $G_5$ corresponds to a de-Sitter stage with the dominance of dark energy. This point is stable for $\alpha<\frac{1}{3}$ and unstable outside this range.

We have sketched the behavior of the critical points $G_3-G_5$ in Fig. \ref{Fig_8}, where we notice that for $\alpha<-\frac{\sqrt{3}+2}{3}$ we have one saddle point and two stable points corresponding to $G_3$ and $G_4$, $G_5$, respectively. In the range $-\frac{\sqrt{3}+2}{3}<\alpha<\frac{\sqrt{3}-2}{3}$, we have no saddle point ($G_3$ and $G_4$ do not exist) and one stable point, $G_5$. For $\frac{\sqrt{3}-2}{3}<\alpha<-\frac{1}{12}$ there is one unstable, one saddle and one stable point, $G_3$, $G_4$ and $G_5$, respectively. In the range $-\frac{1}{12}<\alpha<0$ we have two saddle points, $G_3$ and $G_4$, and one stable point, $G_5$. For $0<\alpha<\frac{1}{3}$ we have one saddle point and one stable point, $G_3$ and $G_5$, respectively. For $\alpha>\frac{1}{3}$ we have two saddle points ($G_3$ and $G_5$) and no stable point.

\begin{figure}[ht]
\centering
\includegraphics[width=0.4\textwidth]{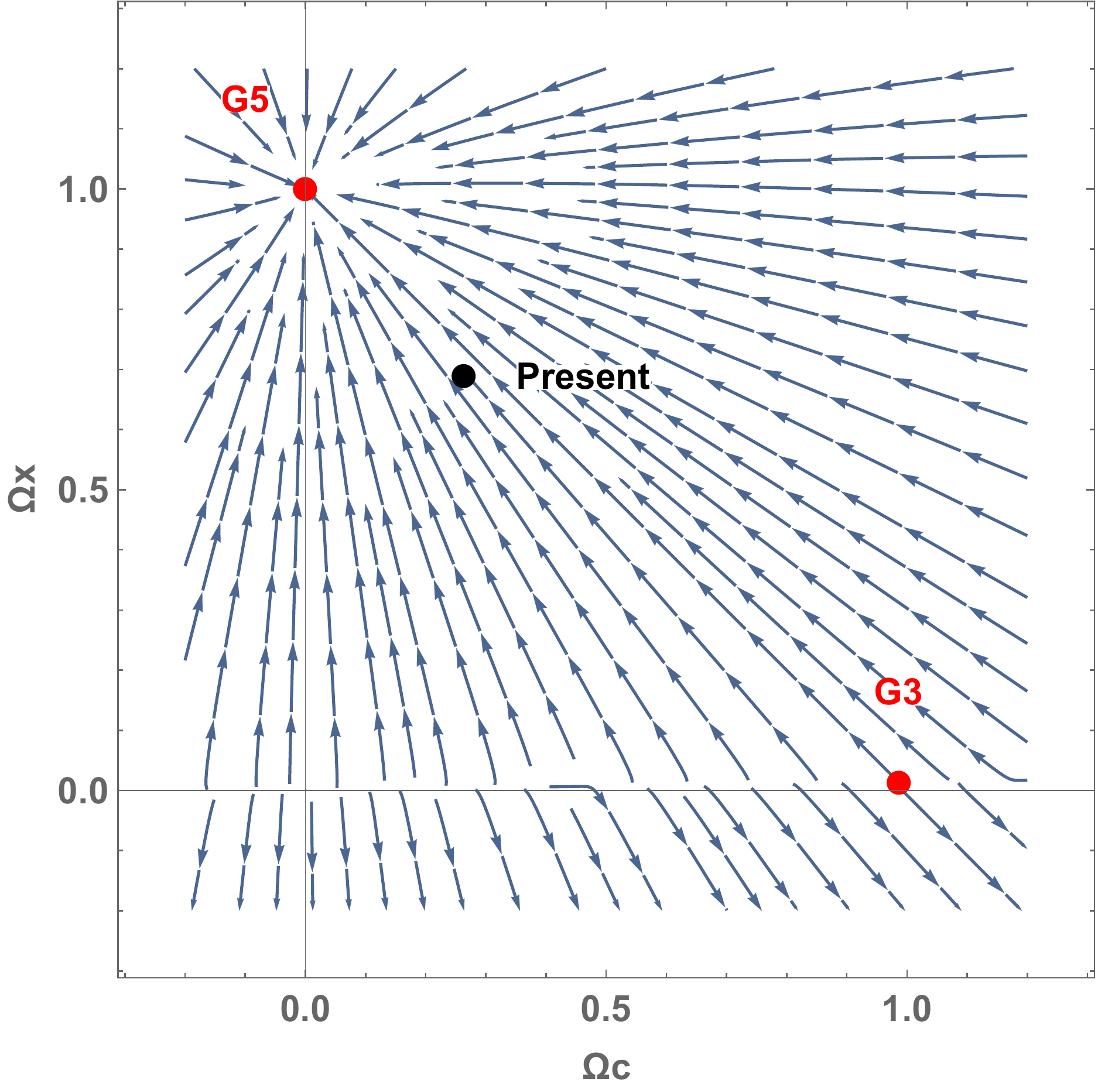}
\caption{Projected phase plots for model $\Gamma_{2c}$ with $\alpha=0.01$ \label{Fig7}}
\end{figure}

\begin{figure*}[ht]
\centering

\includegraphics[width=\textwidth]{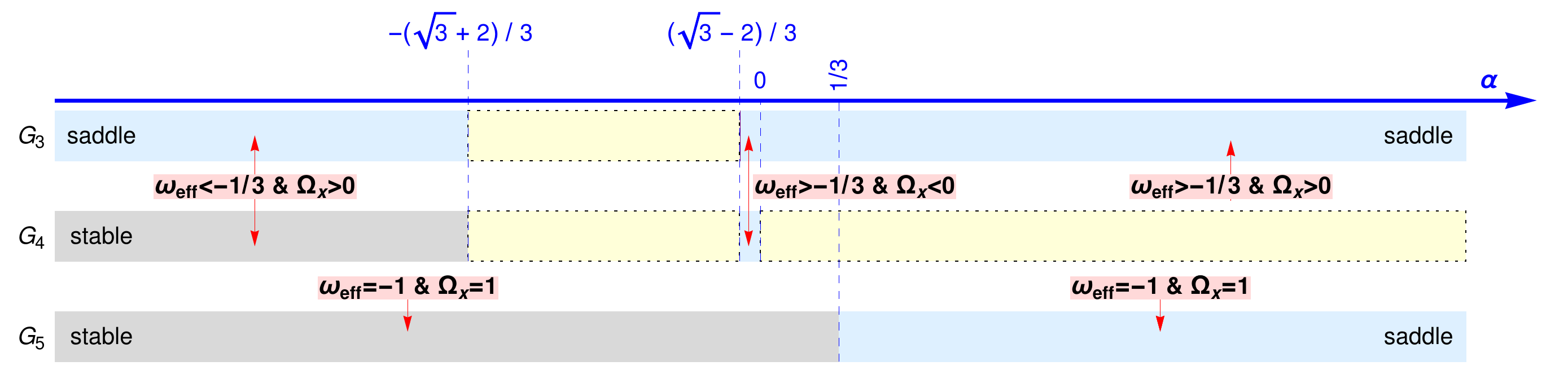}
\caption{Scheme of the behavior of critical points $G_3-G_5$ in terms of the interacting parameter $\alpha$, for model $\Gamma_{2c}$. The pale blue and gray regions represent  a saddle and stable critical point, respectively. The boxed pale yellow regions represent an interval where the existence conditions are not met.
\label{Fig_8}}
\end{figure*}

In Fig. \ref{Fig7} we show the projected 2-dimensional phase space $\Omega_x - \Omega_c$ where we have fixed $\Omega_r=0$ and we have chosen $\alpha=0.01$. In this figure we show two critical points (physical) corresponding to the saddle point $G_3$ and the attractor point $G_5$. The critical point $G_4$ does not exist for $\alpha>0$.

\begin{table*}\begin{center}\begin{minipage}{\textwidth}
\begin{tabular}{|c|c|c|c|}\hline
Critical Points & $\omega_{\rm eff}$ & Existence & Stability Conditions\\\hline
\multirow{2}{*}{$G_1$} & \multirow{2}{*}{$\frac{1}{3}$} &
\multirow{2}{*}{$\alpha\in\mathbb{R}$}& 
{\bf unstable} for $\alpha<-\frac{1}{3}$ or $\alpha>-\frac{1}{12}$\\
& && saddle for $ -\frac{1}{3}<\alpha<-\frac{1}{12}$ \\\cline{1-4}
\multirow{2}{*}{$G_2$} & \multirow{2}{*}{$0$}&
\multirow{2}{*}{$\alpha\in\mathbb{R}$}& 
stable $\alpha=-\frac{1}{6}$ \\ 
& && saddle for $\alpha<-\frac{2}{3}$ or $-\frac{2}{3}<\alpha<-\frac{1}{6}$ or $\alpha > -\frac{1}{6} $\\\cline{1-4}
\multirow{2}{*}{$G_3$} & \multirow{2}{*}{$-\frac{1+6\alpha-\sqrt{1+12\alpha+9\alpha^2}}{9\alpha}$} &
{$\alpha>0$ or $\frac{\sqrt{3}-2}{3}\leq \alpha <0$ }& 
unstable for $\frac{\sqrt{3}-2}{3}<\alpha<-\frac{1}{12}$  \\ 
& &or $\alpha \leq -\frac{\sqrt{3}+2}{3}$&{\bf saddle} for $\alpha<-\frac{\sqrt{3}+2}{3}$ or $\alpha>-\frac{1}{12}$\\\cline{1-4}
\multirow{2}{*}{$G_4$} & \multirow{2}{*}{{$-\frac{1+6\alpha+\sqrt{1+12\alpha+9\alpha^2}}{9\alpha}$}} &
$\frac{\sqrt{3}-2}{3} \le \alpha < 0$& 
{\bf saddle} for $\frac{\sqrt{3}-2}{3}<\alpha<0$   \\\cline{3-4} 
& &{$\alpha \leq -\frac{\sqrt{3}+2}{3}$}&
{\bf stable}  for $\alpha<-\frac{\sqrt{3}+2}{3}$ or $\alpha>0$\\\cline{1-4}
\multirow{2}{*}{$G_5$} & \multirow{2}{*}{-1} &\multirow{2}{*}{$\alpha\in\mathbb{R}$}&  {\bf stable} for $\alpha<\frac{1}{3}$ \\ 
& && saddle for $\alpha>\frac{1}{3}$ \\\hline
\end{tabular}
\end{minipage}
\caption{\label{T8}Description of the critical points for scenario $\Gamma_{2c}$.}
\end{center}
\end{table*}

In summary, if we search for the models having an unstable, saddle and stable critical points during the evolution we get the following intervals, where existence/stability conditions are jointly met: $\alpha>0$ for $\Gamma_{1x}$, $\alpha<0$ for $\Gamma_{1c}$, $\alpha>0$ for $\Gamma_{2d}$ and $0<\alpha<\frac{1}{3}$ for $\Gamma_{2c}$. For these models $\Omega_x>0$ and the sign of the interacting parameter $\alpha$ becomes defined in each case.

On the other hand, the corresponding intervals where existence/stability conditions are jointly met for models where $\Omega_x<0$ are:  $0<\alpha<\frac{4}{9}$ for $\Gamma_{1T}$, $-\frac{1}{9}<\alpha<0$ for $\Gamma_{2T}$ and $0<\alpha<\frac{4}{9}$ for $\Gamma_{1d}$. Models $\Gamma_{1T}$ and $\Gamma_{2T}$ have $\Omega_x<0$ at radiation and dark matter domination, nevertheless $\Omega_x$ changes sign and it becomes positive at dark energy domination. For model $\Gamma_{1d}$ we have $\Omega_x<0$ for dark matter domination, but it is positive for radiation and dark energy domination, meaning at least two change of sign of $\Omega_x$ during evolution.

Finally, notice that the stability conditions of center type critical points have not been addressed in this work. 
\section{Observational analysis}\label{data}

In the observational analysis we use background data such as the local measurement of Hubble parameter ($H_0$)~\cite{Riess:2019cxk}, cosmic chronometers~\cite{Moresco:2016mzx}, supernovae type Ia (SNe Ia)~\cite{Scolnic:2017caz}, baryon acoustic oscillations (BAO)~\cite{Eisenstein:2005su}, and the angular scale of the sound horizon at the last scattering~\cite{Ade:2015rim}. In this section we briefly describe the used data, a detailed description can be find in Ref. \cite{Cid:2018ugy}.

\subsection{Local determination of $H_0$}
We use the $H_0$ value obtained by the SH0ES team using a local distance ladder method based on Cepheids, $h = 0.7402 \pm 0.0142$~\cite{Riess:2019cxk}. 

\subsection{Cosmic chronometers} 
We use 24 cosmic chronometers obtained through the differential age method~\cite{Jimenez:2001gg}. This procedure provides cosmological-independent direct measurements of the expansion history of the universe up to redshift 1.2~\cite{Verde:2014qea}, see Table 3 in Ref. \cite{Cid:2018ugy}.

\subsection{Supernovae Type Ia}
We use the Pantheon Sample, a set of 1048 spectroscopically confirmed SNe Ia ranging from redshift $0.01$ to $2.3$~\cite{Scolnic:2017caz}, along with the corresponding covariance matrix. The Pantheon catalog contains measurements of peak magnitudes in the B-band's rest frame, $m_B$, related to the distance modulus by $\mu=m_B+M_B$. $M_B$ is the absolute B-band magnitude of a fiducial SN Ia, a nuisance parameter.
In our analysis the distance modulus at a given redshift is,
\begin{equation}\label{modulo_de_distancia}
\mu(z)=5\log d_L(z)+25,
\end{equation}
where the luminosity distance $d_L$ is given by,
\begin{equation}
d_L=(1+z)\int^z_0\frac{H_0\, dz'}{H(z')}\, {\rm [Mpc]}.
\end{equation}

\subsection{BAO data}
We use isotropic measurements of the BAO signal from 6dFGS~\cite{Beutler:2011hx}, MGS~\cite{Ross:2014qpa} and eBOSS~\cite{Ata:2017dya}. These measurements are given in terms of the dimensionless ratio,
\begin{equation}
d_z(z) = D_V (z)/r_s(z_d),
\end{equation}
where $z_d$ is the redshift at the drag epoch,
\begin{eqnarray}
D_V(z)&=&\left((1 + z)^2D_A(z)^2 \frac{c\,z}{H(z)}\right)^{1/3}, \\
D_A(z)&=& \frac{c}{(1+z)}\int^z_0 \frac{dz}{H(z)},\\ r_s(z)&=&\int^{\infty}_z \frac{c}{\sqrt{3(1+R)}}\frac{ dz}{H(z)},
\end{eqnarray}
with $c$ the speed of light and $R = \frac{3\Omega_b}{4\Omega_{\gamma}(1+z)}$~\cite{Eisenstein:1997ik}. 

Furthermore, we use the anisotropic BAO measurements from BOSS DR12~\cite{Alam:2016hwk} and Ly$\alpha$ forest~\cite{duMasdesBourboux:2017mrl}, which are defined in terms of $D_A$ and $D_H=c/H(z)$, as shown in table 2 of Ref.~\cite{Evslin:2017qdn}. We use these data along with the corresponding covariance matrix in Ref.~\cite{Evslin:2017qdn}.
\subsection{Cosmic microwave background data}
We consider the angular scale of the sound horizon at the last scattering as the only contribution of CMB data,
\begin{equation}
\ell_a =\frac{\pi (1+z_*)D_A(z_*)}{r_s(z_*)},
\end{equation}
where the comoving size of the sound horizon is evaluated at $z_* = 1089.80$~\cite{Aghanim:2018eyx}. We compare the value obtained in our study with the one reported by the Planck collaboration in 2015, $\ell_a=301.63\pm0.15$~\cite{Ade:2015rim}. 

To compute the maximum likelihood and posterior distributions we use the \textsc{MultiNest} algorithm\footnote{\url{https://github.com/JohannesBuchner/MultiNest}} \cite{Feroz:2007kg,Feroz:2008xx}, with a global log-evidence tolerance of 0.01 as a convergence criterion and working with 800 live points to improve the accuracy.
\vspace{0.5cm}

\begin{figure}[ht]
\centering
\includegraphics[width=0.45\textwidth]{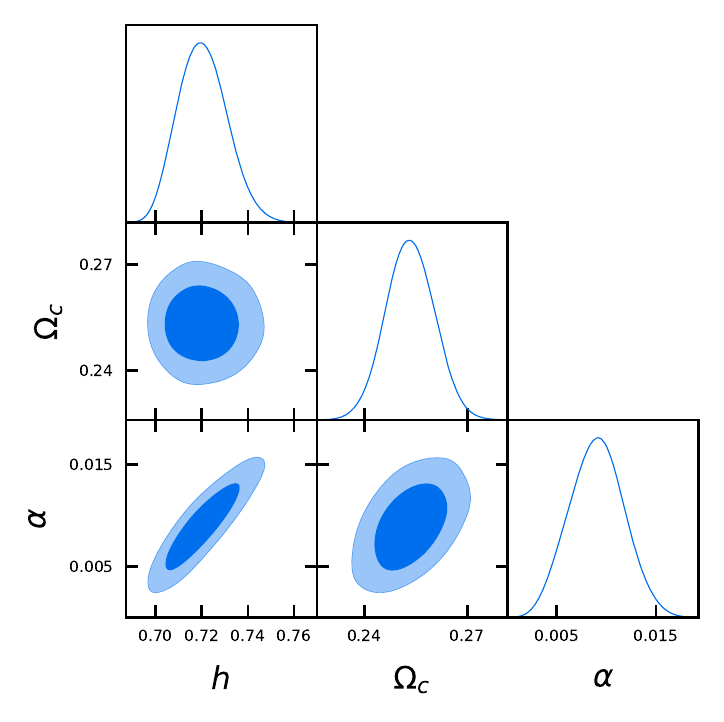}
\caption{Contour plots for the model $\Gamma_{1T}$, $1\sigma$ and $2\sigma$ confidence levels are shown.
\label{Fig8}}
\end{figure}

We perform the observational analysis for the only model with an analytical solution, i.e. scenario $\Gamma_{1T}$. The posterior distribution for the parameters $h$, $\Omega_c$ and $\alpha$ are shown in Fig. \ref{Fig8} and Table \ref{tab_obs}. 

\begin{table*}[ht!]
\centering
\begin{tabular}{|c|c|c|c|}\hline
$h$ & $\Omega_c$  & $\alpha$ & $\chi^2_{\rm min}$  \\ \hline
$0.7204^{+0.0098}_{-0.011}$ & $0.2533\pm0.0070$ & $0.0090\pm0.0027$ & $1044.278$   \\ \hline
\end{tabular}
\caption{\label{tab_obs}Best fit parameters and $1\sigma$ error for the observational analysis of the $\Gamma_{1T}$ scenario.}
\end{table*}

In this analysis we fix the physical baryon density to $\Omega_bh^2 = 0.022383$~\cite{Aghanim:2018eyx}, the physical photon density to $\Omega_{\gamma}h^2=2.469\times10^{-5}$ \cite{Komatsu:2010fb}, we consider the the radiation density as $\Omega_{r}=\Omega_{\gamma}\left(1+\frac{7}{8}\left(\frac{4}{11}\right)^{4/3}N_{\rm eff}\right)$, and the effective number of neutrinos $N_{\rm eff}=3.046$ \cite{Lesgourgues:2018ncw}. Notice that we have considered a flat prior for the $\alpha$ parameter consistent with the results of section \ref{G1T}, i.e., $0<\alpha<\frac{4}{9}$.

\begin{figure*}[ht]
\centering
\includegraphics[width=0.8\textwidth]{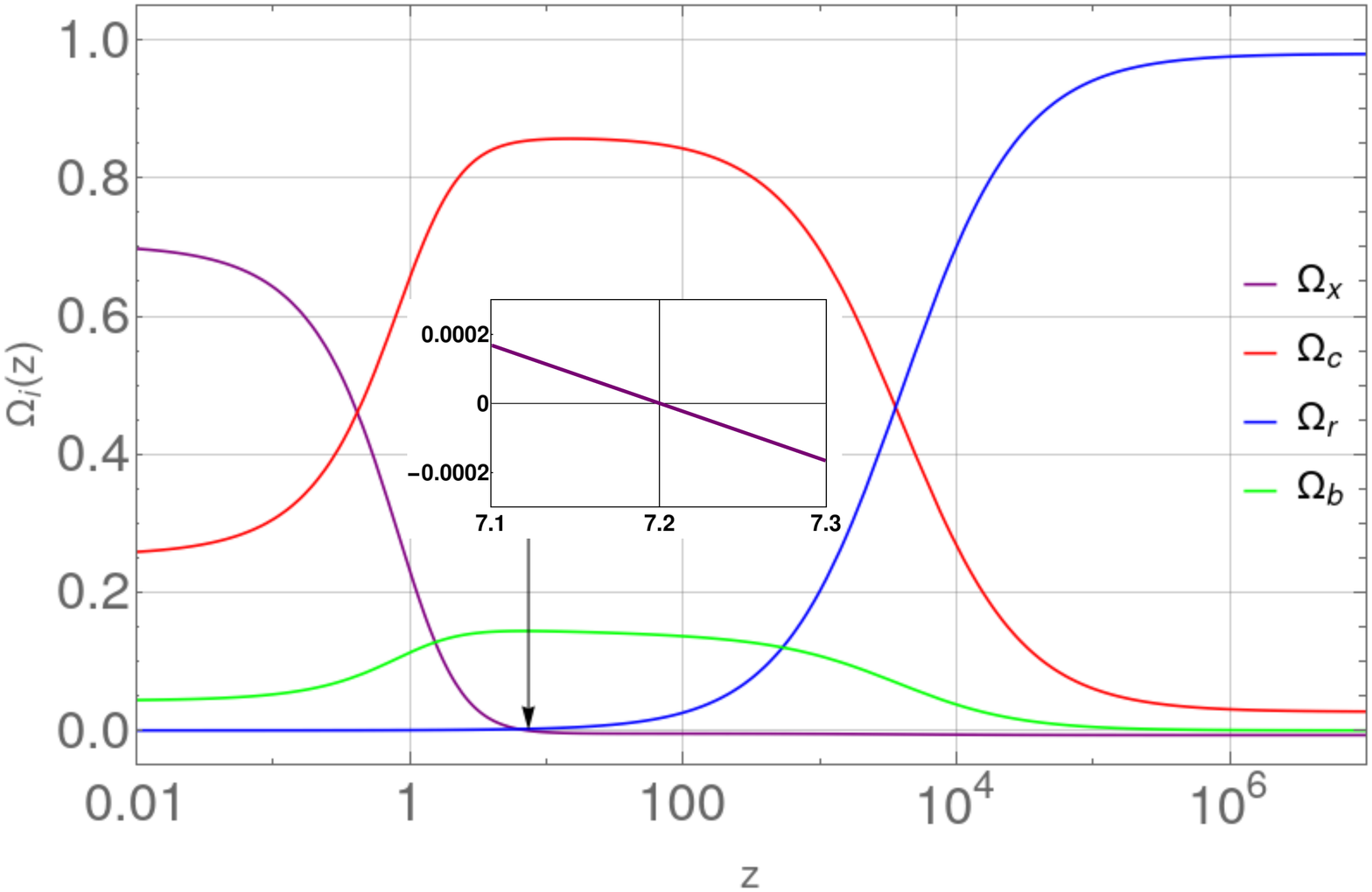}
\caption{Evolution of the density parameters for the model $\Gamma_{1T}$ considering the best fit parameters in Table \ref{tab_obs}.\label{Fig9a}}
\end{figure*}

The results in Table \ref{tab_obs} are consistent with a small but non-null interaction in the dark sector. Given that the interaction $\Gamma_{1T}$ changes sign during evolution, our results indicate that today dark energy is transferred to dark matter but in the past the transference was the opposite. In Fig.\ref{Fig9a} we show the evolution of the density parameters consistent with the best fit parameters shown in Table \ref{tab_obs}. Here we notice that the dark energy density was negative in the past and around $z=7.2$ it transitioned to a positive dark energy density, as shown in the subplot in Fig. \ref{Fig9a}. In the limit $z\rightarrow\infty$, the dark energy density tends to $\Omega_x\rightarrow-0.00675$ instead of 0, as in the $\Lambda$CDM model. Finally, notice that a negative dark energy density in the past is a intrinsic feature of model $\Gamma_{1T}$, which arise from imposing the existence conditions in the dynamical system analysis.

\section{Statefinder parameters \label{SFP}}
The statefinder parameters $r$ and $s$ are introduced in order to compare the plethora of dark energy models available. In the definition of these parameters, derivatives of the scale factor exceed the second-order. 
We can explore the phase space by defining:
\begin{eqnarray}
r \equiv \frac{\dddot a}{a H^3}, \quad s \equiv \frac{r-1}{3\left(q-\frac{1}{2}\right)}.
\end{eqnarray}
In terms of the density parameters $r$ and $s$ are given by: 
\begin{eqnarray}
r&=&\left( 2+\Omega_r-3\Omega_x\right)q-q', \\
s&=&\frac{2\left( -1+q(2+\Omega_r-3\Omega_x)-q'\right)}{3 (\Omega_r-3\Omega_x)},
\end{eqnarray}
where $q$ is defined in (\ref{qa}). 
For each model in Table \ref{Def_int} we can obtain the parameters $r$ and $s$ and the $r-q$ and $r-s$ planes can be drawn. In order to do this we have to numerically integrate equations \eqref{eq1}-\eqref{eq3} and we have considered as border conditions the present values of the density parameters as $\{\Omega_m,\Omega_c,\Omega_x\}=\{0.3111,0.26212,0.6889\}$ \cite{Planck:2018vyg}. 
The trajectories in the  planes $r-q$ and $r-s$  can exhibit quite different behaviors, even though the overall behavior of models may appear similar in the dynamical system analysis of the previous section. 
\begin{figure*}[ht]
\centering
\includegraphics[width=0.45\textwidth]{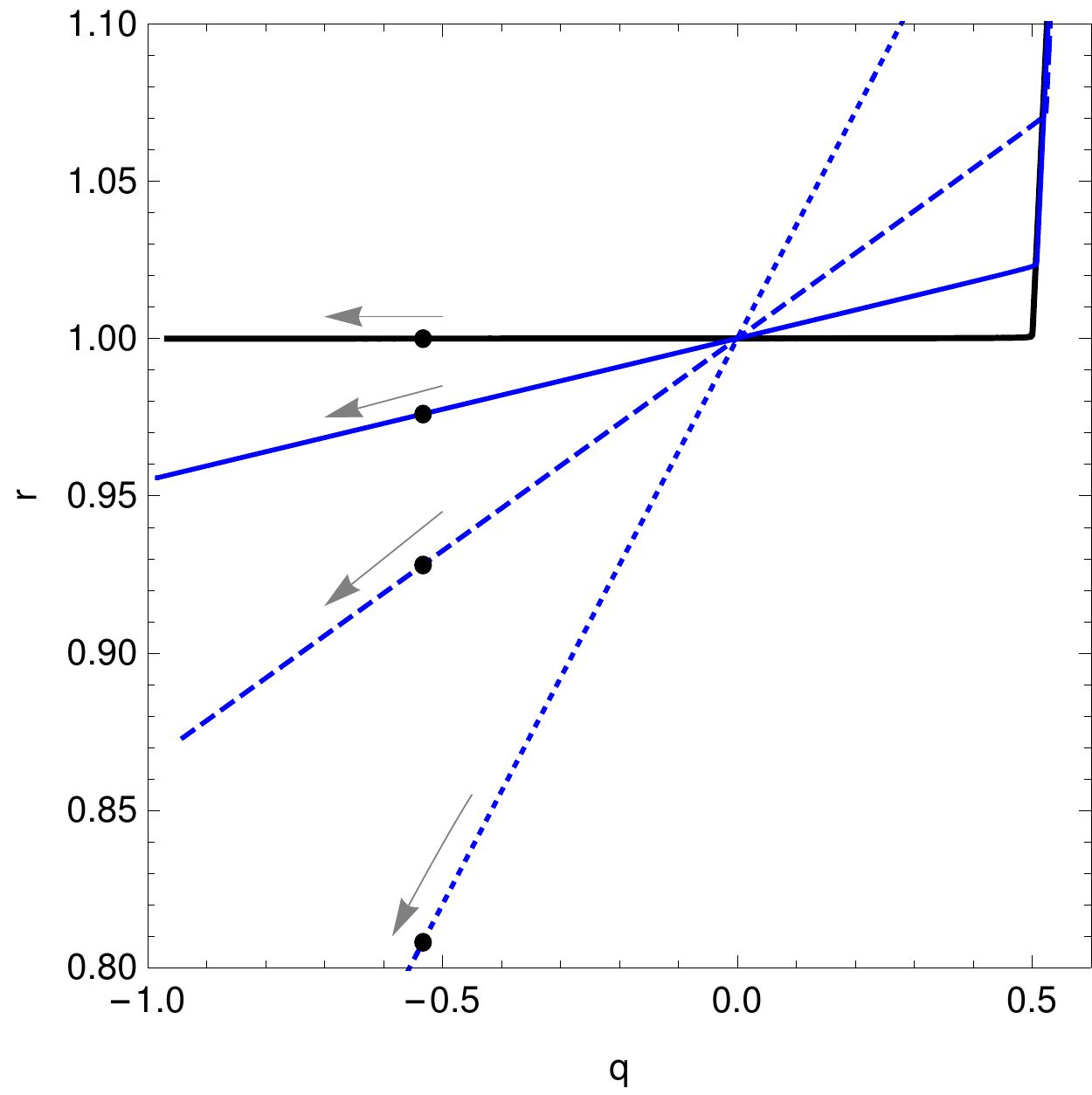}
\includegraphics[width=0.45\textwidth]{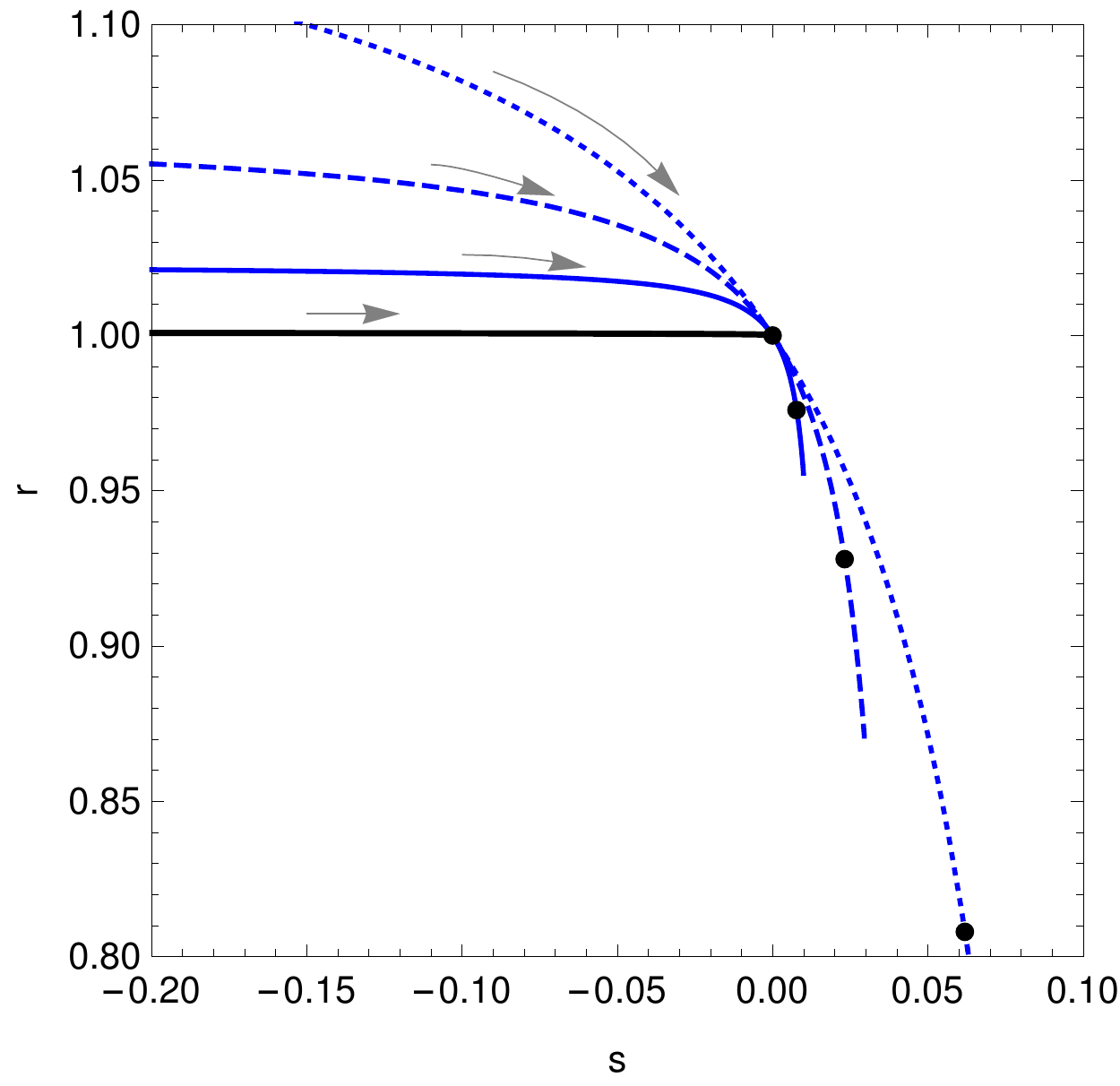}
\caption{The figures show the space parameter $r-q$ (left) and $r-s$ (right) for the $\Gamma_{1T}$ model and several values of the $\alpha $ parameter. Black dots indicate current values and gray arrows the evolution's direction. The solid black line represents $\Lambda$CDM, meanwhile, blue solid, dashed and dotted lines indicate $\alpha=0.01,0.03,0.08$, respectively.
\label{Fig9}}
\end{figure*}

\begin{figure*}[ht]
\centering
\includegraphics[width=0.45\textwidth]{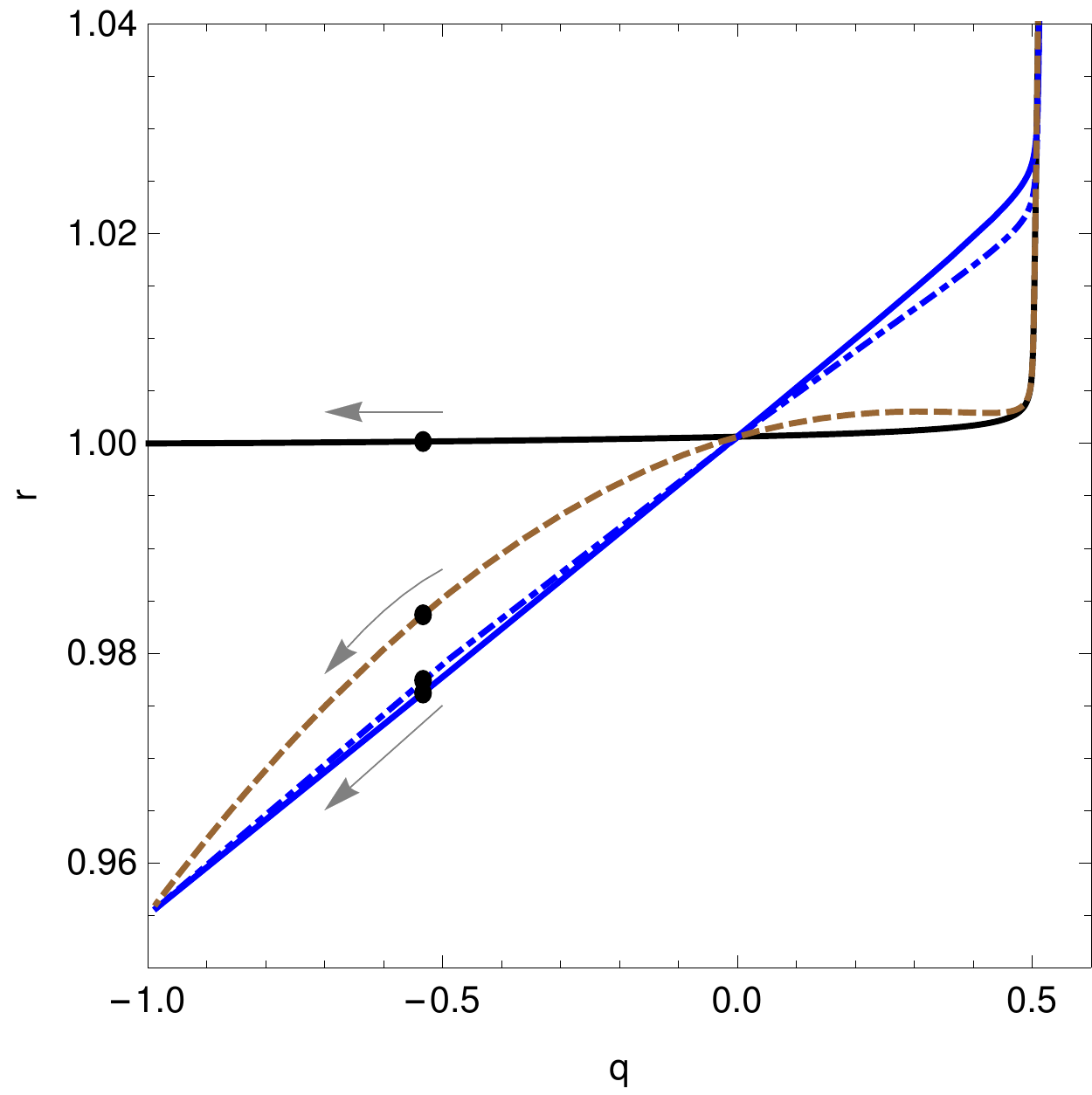}
\includegraphics[width=0.45\textwidth]{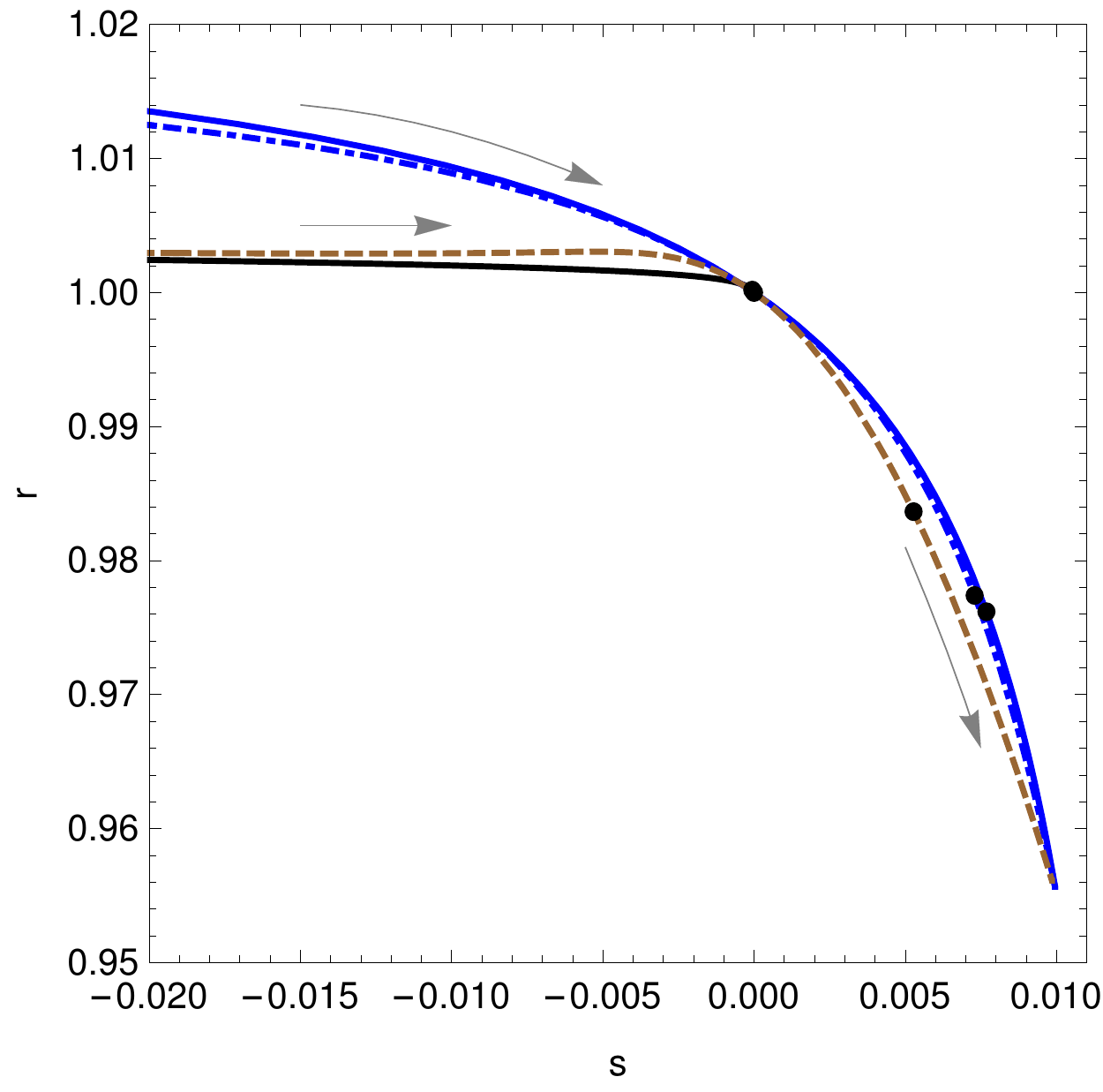}
\caption{The figure shows the planes $r-q$ (left) and $r-s$ (right) for the $\Gamma_{1T}$, $\Gamma_{1d}$ and $\Gamma_{1x}$ models in the case $\alpha=0.01$. Black dots indicate current values and gray arrows the evolution's direction. The solid black line represents $\Lambda$CDM, meanwhile, blue solid, blue dot-dashed and brown lines indicate $\Gamma_{1T}$, $\Gamma_{1d}$ and $\Gamma_{1x}$, respectively.
\label{Fig10}}
\end{figure*}

\begin{figure*}[ht]
\centering
\includegraphics[width=0.45\textwidth]{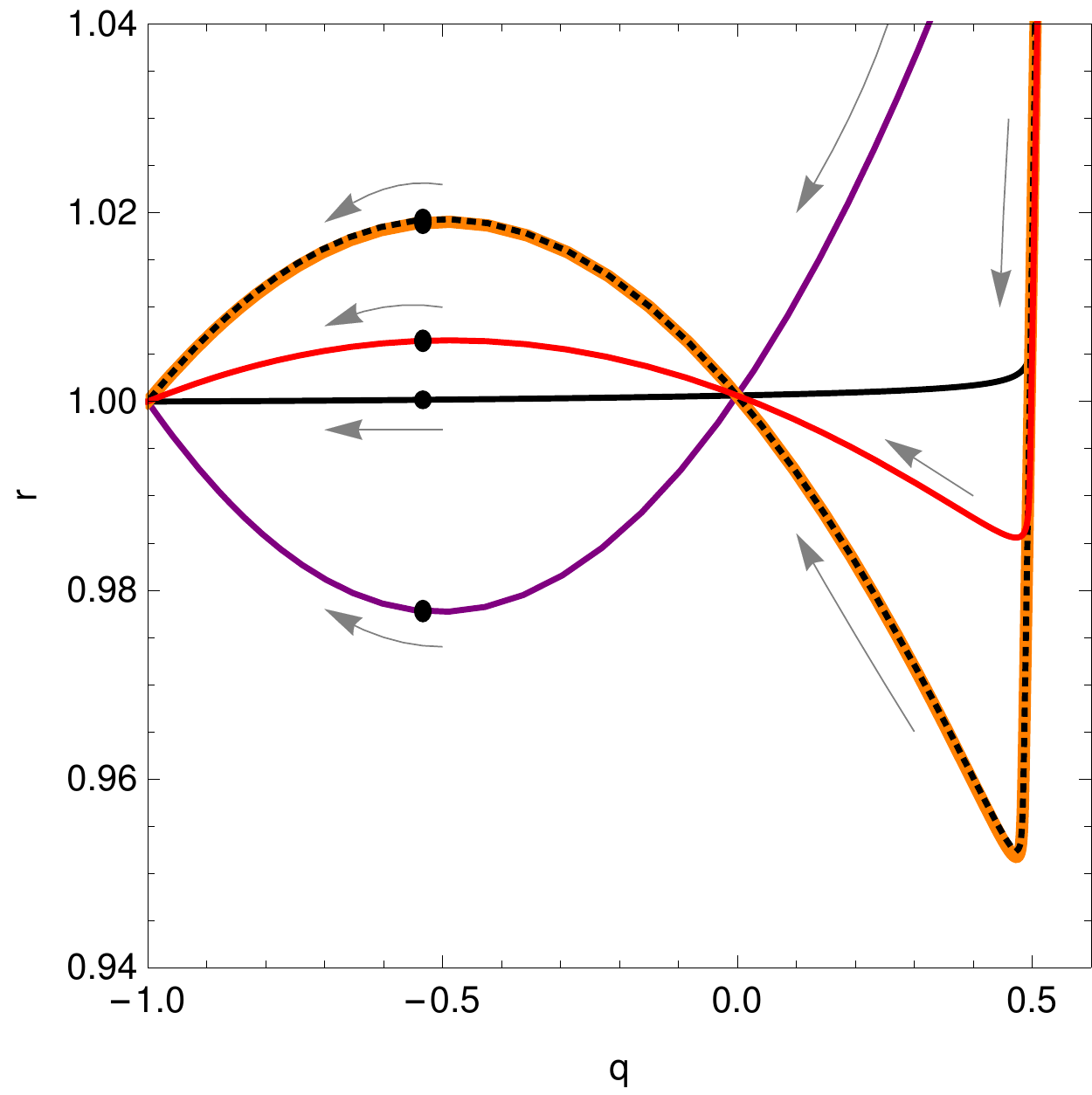}
\includegraphics[width=0.45\textwidth]{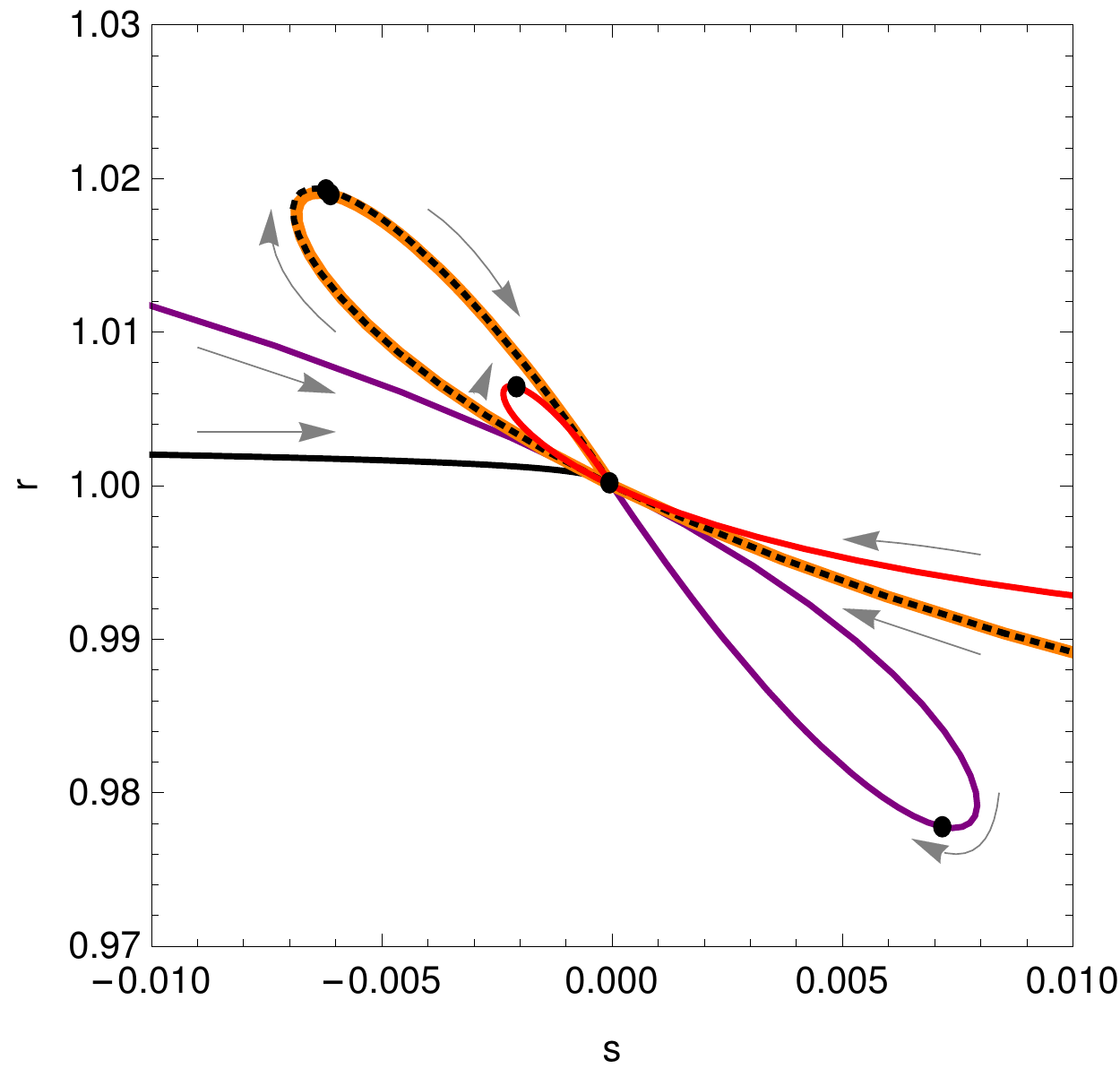}
\caption{The figure shows the planes $r-q$ (left) and $r-s$ (right) for the models $\Gamma_{2T}$ and $\Gamma_{1c}$ in the case $\alpha=-0.01$ and models $\Gamma_{2c}$ and $\Gamma_{2d}$ in the case $\alpha=0.01$. Black dots indicate current values and gray arrows the evolution's direction. The solid black line represents $\Lambda$CDM, meanwhile, purple, red, orange and black dotted lines correspond to models $\Gamma_{2T}$, $\Gamma_{1c}$, $\Gamma_{2d}$ and $\Gamma_{2c}$, respectively.
\label{Figura11}}
\end{figure*}

The $r-q$ and $r-s$ planes for the $\Gamma_{1T}$ model are shown in Fig. \ref{Fig9}, here the trajectories correspond to different values of $\alpha$ consistent with the results of section \ref{G1T}. We notice that in both planes the variation in $\alpha$ modify the the trajectories with respect to the $\Lambda$CDM model.

In Fig. \ref{Fig10} we show the  $r-q$ and $r-s$ planes for  the models $\Gamma_{1T}$, $\Gamma_{1d}$ and $\Gamma_{1x}$ in the case $\alpha=0.01$.  From the  trajectories we see that all the curves pass through the points $(r,q)=(1,0)$ and $(r,s)=(1,0)$, and they present the same outcomes in the future. Notice that models $\Gamma_{1T}$ and $\Gamma_{1d}$ are very similar in both planes.

Fig. \ref{Figura11} shows the $r-q$ and $r-s$ planes for models $\Gamma_{1c}$, $\Gamma_{2T}$, $\Gamma_{2d}$ and $\Gamma_{2c}$, we consider $\alpha=-0.01$ for models $\Gamma_{2T}$ and $\Gamma_{2d}$ and $\alpha=0.01$ for models $\Gamma_{2d}$ and $\Gamma_{2c}$, consistent with the results in section \ref{DS1}. We see that all trajectories in Figure \ref{Figura11} pass through the points $(r,q)=(1,0)$ and they end at the points $(r,q)=(1,-1)$ and $(r,s)=(1,0)$. Even though the interacting models trajectories end in the same point as $\Lambda$CDM in the plane $r-s$, the interacting models present a loop in their trajectories, passing through the point $(r,s)=(1,0)$ twice. 
 
We notice that the ending point is different from $\Lambda$CDM for Figures \ref{Fig9} and \ref{Fig10}, but is the same point for Fig. \ref{Figura11}. The trajectories in Fig. \ref{Fig10} tend to an asymptotic point corresponding to a scaling solution and in Fig. \ref{Figura11} the trajectories evolve towards the de Sitter point. Notice that the trajectories in Fig. \ref{Figura11} have different shapes from those shown in Fig. \ref{Fig10}, this separation seems not to be related with the classification of models in \eqref{Int_i}-\eqref{Int_j}.

Finally, similar trajectories to those shown in Figures \ref{Fig10} and \ref{Figura11} are found in Ref. \cite{Panotopoulos:2020kpo}, where two different interacting models where studied, which do not have a change of sign in the interacting term.

\section{Final Remarks\label{FR}}
We have performed a dynamical system analysis and a statefinder analysis for a class of sign-changeable interactions in the dark sector \cite{Arevalo:2019axj}. 
We have shown that the studied models can explain the current acceleration of our universe as a late time attractor in a defined range for the $\alpha$ parameter (see section \ref{DS1}). For the studied models the evolution of the universe can go from a radiation dominated era, through a dark matter dominated era and a final stage of dark energy domination. These kind of evolution is found in each model for a defined range of the interaction parameter $\alpha$ (see section \ref{DS1}). We find that models, $\Gamma_{1T}$, $\Gamma_{2T}$, $\Gamma_{1d}$, need to have $\Omega_x<0$ at some critical points in order to follow the universe's evolution, this necessarily implies at least one change of sign of $\Omega_x$ during evolution, resulting in $\Omega_x>0$ at the final attractor. On the other hand, the requirement of the existence and stability conditions for the critical points selects in each case a defined sign for the interacting parameter $\alpha$.

On the other hand, the observational analysis of model $\Gamma_{1T}$ with background data shows that the posterior distribution of the $\alpha$ parameter is  consistent with the result from the dynamical system analysis. Besides, this model includes a change of sign for the dark energy density parameter, which must have been negative in the past.

In order to distinguish interacting models from $\Lambda$CDM, we apply the statefinder analysis. The statefinder diagrams show that for some of the models, $\Gamma_{2T}$, $\Gamma_{1c}$, $\Gamma_{2d}$ and $\Gamma_{2c}$,
the trajectories will eventually
approach to the $\Lambda$CDM fixed point in the $r-s$ and $r-q$ planes (see Fig.\ref{Figura11}). Nevertheless, models that asymptotically tend to a scaling state, $\Gamma_{1T}$, $\Gamma_{1d}$ and $\Gamma_{1x}$,  do not follow this lineament and have a different asymptotic behavior than $\Lambda$CDM in the  $r-q$ and $r-s$ planes (see Fig.\ref{Fig10}).
Furthermore, the statefinder diagrams indicate that the interaction has a significant effect on the evolution of the universe, and the interacting models can be distinguished from the $\Lambda$CDM model by this analysis.
\vspace{0.5cm}

{\it Acknowledgements.} AC acknowledges the support of Vicerrectoría de Investigación y Postgrado of Universidad del Bío-Bío through grant no. 2120247 IF/R.

\end{document}